\documentclass[useAMS,usenatbib]{mn2e}
\usepackage{psfrag,graphicx}
\usepackage{color}
\usepackage{txfonts}
\usepackage{breqn}
\usepackage[T1]{fontenc}
\usepackage{ae,aecompl}
\title[Black hole formation in the early universe]
{A UV flux constraint on the formation of direct collapse black holes}
\author[Latif et al.]
  {M. A. Latif, \thanks{Corresponding author: mlatif@astro.physik.uni-goettingen.de}$^1$
  S. Bovino,$^1$
  C. Van Borm, $^{1,2}$
  T. Grassi, $^{3,4}$
 \newauthor
  D. R. G. Schleicher, $^1$
  M. Spaans$^2$\\
   $^1$ Institut f\"ur Astrophysik, Georg-August-Universit\"at, Friedrich-Hund-Platz 1, D-37077 G\"ottingen, Germany \\
   $^2$ Kapteyn Astronomical Institute, University of Groningen, The Netherlands  \\
   $^3$Centre for Star and Planet Formation, Natural History Museum of Denmark, \O ster Voldgade 5-7, DK-1350 Copenhagen, Denmark\\
   $^4$Niels Bohr Institute, University of Copenhagen, Juliane Maries Vej 30, DK-2100 Copenhagen, Denmark 
  }
\date{}

\pagerange{\pageref{firstpage}--\pageref{lastpage}} \pubyear{2009}

\def\LaTeX{L\kern-.36em\raise.3ex\hbox{a}\kern-.15em
      T\kern-.1667em\lower.7ex\hbox{E}\kern-.125emX}

\begin{document}

\bibliographystyle{mn2e}

\label{firstpage}

\maketitle
\def\na{NewA}%
\def\aj{AJ}%
\def\araa{ARA\&A}%
\def\apj{ApJ}%
\def\apjl{ApJ}%
\def\jcap{JCAP}

\def\apjs{ApJS}%
\def\ao{Appl.~Opt.}%
\def\apss{Ap\&SS}%
\def\aap{A\&A}%
\def\aapr{A\&A~Rev.}%
\def\aaps{A\&AS}%
\def\azh{AZh}%
\def\baas{BAAS}%
\def\jrasc{JRASC}%
\def\memras{MmRAS}%
\def\mnras{MNRAS}%
\def\pra{Phys.~Rev.~A}%
\def\prb{Phys.~Rev.~B}%
\def\prc{Phys.~Rev.~C}%
\def\prd{Phys.~Rev.~D}%
\def\pre{Phys.~Rev.~E}%
\def\prl{Phys.~Rev.~Lett.}%
\def\pasp{PASP}%
\def\pasj{PASJ}%
\def\qjras{QJRAS}%
\def\skytel{S\&T}%
\def\solphys{Sol.~Phys.}%

\def\sovast{Soviet~Ast.}%
\def\ssr{Space~Sci.~Rev.}%
\def\zap{ZAp}%
\def\nat{Nature}%
\def\iaucirc{IAU~Circ.}%
\def\aplett{Astrophys.~Lett.}%
\def\apspr{Astrophys.~Space~Phys.~Res.}%
\def\bain{Bull.~Astron.~Inst.~Netherlands}%
\def\fcp{Fund.~Cosmic~Phys.}%
\def\gca{Geochim.~Cosmochim.~Acta}%
\def\grl{Geophys.~Res.~Lett.}%
\def\jcp{J.~Chem.~Phys.}%
\def\jgr{J.~Geophys.~Res.}%
\def\jqsrt{J.~Quant.~Spec.~Radiat.~Transf.}%
\def\memsai{Mem.~Soc.~Astron.~Italiana}%
\def\nphysa{Nucl.~Phys.~A}%
\def\physrep{Phys.~Rep.}%
\def\physscr{Phys.~Scr}%
\def\planss{Planet.~Space~Sci.}%
\def\procspie{Proc.~SPIE}%

%


\begin{abstract}
{
The ability of metal free gas to cool by molecular hydrogen in primordial halos is strongly associated with the strength of ultraviolet (UV) flux produced by the stellar populations in the first galaxies. Depending on the stellar spectrum, these UV photons can either dissociate $\rm H_{2}$ molecules directly or indirectly by photo-detachment of $\rm H^{-}$ as the latter provides the main pathway for $\rm H_{2}$ formation in the early universe. In this study, we aim to determine the critical strength of the UV flux above which the formation of molecular hydrogen remains suppressed for a  sample of five distinct halos at $z>10$ by employing a higher order chemical solver and a Jeans resolution of 32 cells. We presume that such flux is emitted by PopII stars implying atmospheric temperatures of $\rm 10^{4}$~K. We performed three-dimensional cosmological simulations and varied the strength of the UV flux below the Lyman limit in units of $\rm J_{21}$. Our findings show that the value of $\rm J_{21}^{crit}$ varies from halo to halo and is sensitive to the local thermal conditions of the gas. For the simulated halos it varies from 400-700 with the exception of one halo where $\rm J_{21}^{crit} \geq 1500$. This has important implications for the formation of direct collapse black holes and their estimated population at z > 6. It reduces the number density of direct collapse black holes by almost three orders of magnitude compared to the previous estimates. 
}
\end{abstract}


\begin{keywords}
methods: numerical -- cosmology: theory -- early Universe -- galaxies: formation
\end{keywords}

\section{Introduction}

Observations of quasars at z $>6$ reveal that supermassive black holes (SMBHs) of a few billion solar masses were assembled within the first billion years after the Big Bang \citep{2003AJ....125.1649F,2006AJ....131.1203F,2010AJ....139..906W,2011Natur.474..616M,2013ApJ...779...24V}. Their formation mechanisms in the juvenile Universe remain unknown. The potential progenitors of SMBHs include the remnants of Pop III stars \citep{2001ApJ...552..459H,2004ApJ...613...36H,2009ApJ...696.1798T,2012ApJ...756L..19W,2014ApJ...781...60H,2014ApJ...784L..38M}, dense stellar cluster \citep{2004Natur.428..724P,2008ApJ...686..801O, 2009ApJ...694..302D} and direct collapse of a protogalactic gas cloud into so-called direct collapse black holes (DCBHs) \citep{2002ApJ...569..558O,2003ApJ...596...34B,2006ApJ...652..902S,2006MNRAS.370..289B,2006MNRAS.371.1813L,2008MNRAS.391.1961D,2008arXiv0803.2862D,2010MNRAS.402.1249S,2010MNRAS.tmp.1427J,2010A&ARv..18..279V,2010ApJ...712L..69S,2011MNRAS.411.1659L,2012RPPh...75l4901V,2012arXiv1203.6075H,2013MNRAS.436.2301P,2013MNRAS.433.1607L,2013MNRAS.tmp.2526L,2013ApJ...774...64W,2013ApJ...771...50A,2013MNRAS.433.1556Y,2013MNRAS.436.2989L,2013A&A...560A..34W,2014MNRAS.tmp..564L,2014MNRAS.440.1263Y,2014MNRAS.440.2969L,2014arXiv1404.4630I,2014arXiv1403.1293V}.

Pristine massive primordial halos of $\rm 10^{7}-10^{8}~M_{\odot}$ which formed in the early universe at z$=15-20$ are the potential cradles for these DCBHs. It is imperative that their halos remain metal free and cooling is mainly regulated by atomic line radiation instead of $\rm H_{2}$. These conditions may lead to a monolithic isothermal collapse where fragmentation is suppressed and a supermassive star of $\rm 10^{4}-10^{6}~M_{\odot}$ forms which later collapses into a black hole (i.e., DCBH). This scenario is supported by numerical simulations which show that fragmentation remains inhibited and massive objects may form \citep{2003ApJ...596...34B,2008ApJ...682..745W,2009MNRAS.393..858R,2011MNRAS.411.1659L,2013MNRAS.432..668L,2013MNRAS.430..588L,2013MNRAS.433.1607L}. Recently, the feasibility of this scenario has been explored via high resolution numerical experiments and it is found that $\rm \sim10^{5}~M_{\odot}$ objects can form \citep{2013MNRAS.436.2989L,2014MNRAS.439.1160R}. 
These results are consistent with theoretical predictions \citep{2008MNRAS.387.1649B,2010MNRAS.402..673B,2011MNRAS.414.2751B,2012ApJ...756...93H,2012MNRAS.421.2713B,2013A&A...558A..59S,2013ApJ...778..178H,2013ApJ...768..195W}. Depending on the mass accretion rates these studies suggest the formation of supermassive stars or quasi-stars (stars with BH at their center) as potential embryos of DCBHs \citep{2013A&A...558A..59S}.

In primordial gas, trace amount of $\rm H_{2}$ can be formed via gas phase reactions in the early universe which then leads to cooling and star formation. The main channel for the formation of $\rm H_{2}$ is:
\begin{equation}
\mathrm{H + e^{-} \rightarrow H^{-} +} \gamma
\end{equation}
\begin{equation}
\mathrm{ H + H^{-} \rightarrow  H_{2} + e^-.}\\ 
\label{h21}
\end{equation}
Once the first generation of stars, so-called PopIII stars, are formed they produce UV flux, pollute the intergalactic medium with metals via supernova explosions, and lead to a second generation of stars known as PopII stars. The UV flux produced by these stellar populations either photo-dissociates the molecular hydrogen directly or photo-detaches electrons from H$^-$ which provides the main route for the formation of H$_{2}$ in primordial gas chemistry.

The stellar spectra of PopIII stars are harder with a characteristic temperature of $\rm 10^{5}~K$ while PopII stars are characterized by soft spectra with temperatures of $\rm 10^{4}~K$. PopIII stars mainly contribute to the direct dissociation of H$_2$ while PopII stars also photo-detach H$^-$. UV photons with energies between 11.2 and 13.6 eV are absorbed in the Lyman-Werner bands of molecular hydrogen and put it into an excited state. The $\rm H_{2}$ molecule later decays to the ground state and gets dissociated as
\begin{equation}
 \mathrm{ H_{2}} + \gamma  \mathrm{ \rightarrow  H_{2}^{*} \rightarrow  H + H } , \\
\end{equation}
a process known as the Solomon process. On the other hand, H$^-$ photo-detachment occurs via low energy photons above 0.76 eV.  In this study, we focus on the background UV flux predominantly emitted by PopII stars as tiny amounts of metals can lead to fragmentation \citep{2005ApJ...626..627O,2013ApJ...766..103D}.

The critical value of the UV flux, hereafter called $\rm J_{21}^{crit}$, above which $\rm H_{2}$ cooling remains suppressed, can be determined by comparing the $\rm H_{2}$ formation and dissociation time scales. \cite{2001ApJ...546..635O} found from one-zone calculations that $\rm J_{21}^{crit}=10^3$ in units of $\rm J_{21}=10^{-21}~erg~cm^{-2}~s^{-1}~Hz^{-1}~sr^{-1}$ for $\rm T_{*}=10^4~K$ which was later confirmed by \cite{Bromm03} in 3D simulations for a single halo. These estimates were revised by \cite{2010MNRAS.402.1249S} (hereafter S10) through three dimensional simulations using the $\rm H_{2}$ self-shielding formula of \cite{1996ApJ...468..269D}, finding that $\rm J_{21}^{crit}= 30-300$. They attributed these differences to the choice of a more accurate and higher $\rm H_{2}$ collisional dissociation rate, and focused on rather massive halos forming at z $<$10. \cite{2011MNRAS.418..838W} (hereafter  WG11) improved the H$_2$ self-shielding function of \cite{1996ApJ...468..269D}  and anticipated that it may further reduce the value of $\rm J_{21}^{crit}$. Such values 
of $\rm J_{21}^{crit}$ are much larger than the global background flux but can be achieved in the close vicinity (about 10 kpc) of nearby star forming 
galaxies \citep{2008MNRAS.391.1961D,2012MNRAS.425.2854A,2014arXiv1403.5267A}.

In this article, we derive the values of $\rm J_{21}^{crit}$ for a stellar spectrum of $\rm T_{*}=10^4~K$ employing the improved $\rm H_{2}$ self-shielding fitting function provided by \cite{2011MNRAS.418..838W}. Major improvements compared to  the previous studies are the following:
\begin{itemize}
\item Selection of a larger sample of halos with collapse redshifts at  $z>$10.
\item Employed higher order chemical solver DLSODES \citep{2013MNRAS.434L..36B}.
\item Accurate determination of $\rm J_{21}^{crit}$ for the individual halos.
\item Higher Jeans resolution of 32 cells.
\item Improved self-shielding function of WG11.
\end{itemize}
We note that the importance of an accurate chemical solver in high resolution simulations was previously reported by \cite{2013MNRAS.434L..36B}. The impact of higher Jeans resolution has also been shown by \cite{2013MNRAS.430..588L} and \cite{2012ApJ...745..154T}. Our selected halos are collapsed at $z>$10 in contrast to S10 where halos collapsed at $z<$10. All these improvements distinguish the present work from S10.
 
We perform three dimensional cosmological simulations for five different halos of a few times $\rm 10^{7}~M_{\odot}$ and vary the strength of the background UV flux (hereafter  $\rm J_{21}$, i.e. UV flux below Lyman limit). We use the chemical network listed in table A1 of S10 which includes all the relevant process for the formation and dissociation of molecular hydrogen. We further employed a fixed Jeans resolution of 32 cells throughout the simulations for better resolving the shocks and turbulence. A particular goal of this paper is to provide a rather narrow constraint on $\rm J_{21}^{crit}$ for individual halos, and to point at potential correlations with halo properties. This study has important implications for the formation of DCBHs as it provides stronger estimates for the value of $\rm J_{21}^{crit}$ required for dissociation of molecular hydrogen.

The organization of this article is as follows. In section 2, we provide the details of simulations setup and summary of a chemical network. In the third section, we present our findings and discuss our conclusions in section 4.

\section{Computational methods}
The simulations presented here are performed using the adaptive mesh refinement, grid based, cosmological hydrodynamics code  ENZO\footnote{http://enzo-project.org/, changeset:48de94f882d8} \citep{2004astro.ph..3044O,2014ApJS..211...19B}. The hydrodynamical equations are solved employing the piece-wise parabolic method (PPM) and the dark matter dynamics is computed using the particle-mesh technique. The code makes use of multigrid Poison solver for solving the gravity.

Our simulations start at z=100 with cosmological initial conditions. We first run $\rm 128^3$ uniform grid simulations and select the most massive halos of a few times $\rm 10^{7}~M_{\odot}$ in our simulated periodic box by using a standard halo finder based on the friends of friends algorithm \citep{2011ApJS..192....9T}. Our computational volume has a size of 1 $\rm Mpc/h$ and is centered on the most massive halo.  We employ two nested refinement levels each with a grid resolution of $\rm 128^3$ cells besides the top grid resolution of $\rm 128^3$. To solve the dark matter (DM) dynamics, 5767168 particles are used which yield an effective DM resolution of about 600 M$_{\odot}$. Further, additional 18 levels of refinement were employed during the course of the simulations with a fixed Jeans resolution of 32 cells. Our refinement strategy is exactly the same as mentioned in a number of previous studies \citep{2013MNRAS.433.1607L,2013MNRAS.tmp.2526L,2014MNRAS.tmp..564L}. For further details about the simulation setup the reader is referred to the above mentioned articles. The simulations were stopped once they reached the maximum refinement level, and were performed for five distinct halos selected from cosmological initial conditions. The masses of these halos and their collapse redshifts are listed in table \ref{table0} for various values of $\rm J_{21}$.

To self-consistently solve the evolution of the following chemical species  $\rm H$, $\rm H^{+}$, $\rm He$, $\rm He^{+}$,~$\rm He^{++}$, $\rm e^{-}$,~$\rm H^{-}$,~$\rm H_{2}$,~$\rm H_{2}^{+}$ in cosmological simulations, we employed the publicly available KROME package\footnote{Webpage KROME: www.kromepackage.org} \citep{2014MNRAS.439.2386G}. The KROME package was previously employed for 3D simulations of primordial star formation and halo mergers \citep{2013MNRAS.434L..36B,2014MNRAS.441.2181B}. The chemical network used in this study is the same as listed in table A1 of S10 with only two modifications, the inclusion of an improved fitting formula for $\rm H_{2}$ self-shielding by WG11 and an addition of the dissociative tunneling effect which contributes to the collisional dissociation of $\rm H_{2}$ \citep{1996ApJ...461..265M}. The latter is the dominant factor in the total dissociation rate ($\gamma_{tot}= \gamma_{CID}+\gamma_{DT}$) of $\rm H_2$ for temperatures up to 4500 K, see detailed discussion in section 3.3 of \cite{1996ApJ...461..265M}. We presume that the background UV flux is emitted by PopII stars with an atmospheric temperature of $\rm T_{*}=10^4~K$. 

Here, we are interested in the UV flux below the Lyman limit (i.e. $\rm< 13.6~eV$). As pointed out in the previous section, the low energy photons with energy above 0.76 eV emitted by PopII stars photo-detach  $\rm H^{-}$ which provides the main pathway for the formation of $\rm H_{2}$ in primordial composition of the gas. We have not included the photo-ionization of hydrogen and helium species as they require energies above the Lyman limit. The H$_{2}$ cooling function of \cite{1998A&A...335..403G} is used here as in S10 for direct comparison and all other relevant cooling processes for primordial gas are included i.e, cooling by collisional excitation, collisional ionization, recombination and Bremsstrahlung radiation \cite[for details see][]{2014MNRAS.439.2386G}. We checked that using the cooling function by \cite{2008MNRAS.388.1627G} has no impact on the results.

\begin{figure*}
\hspace{-6.0cm}
\centering
\begin{tabular}{c}
\begin{minipage}{6cm}
\includegraphics[scale=1.0]{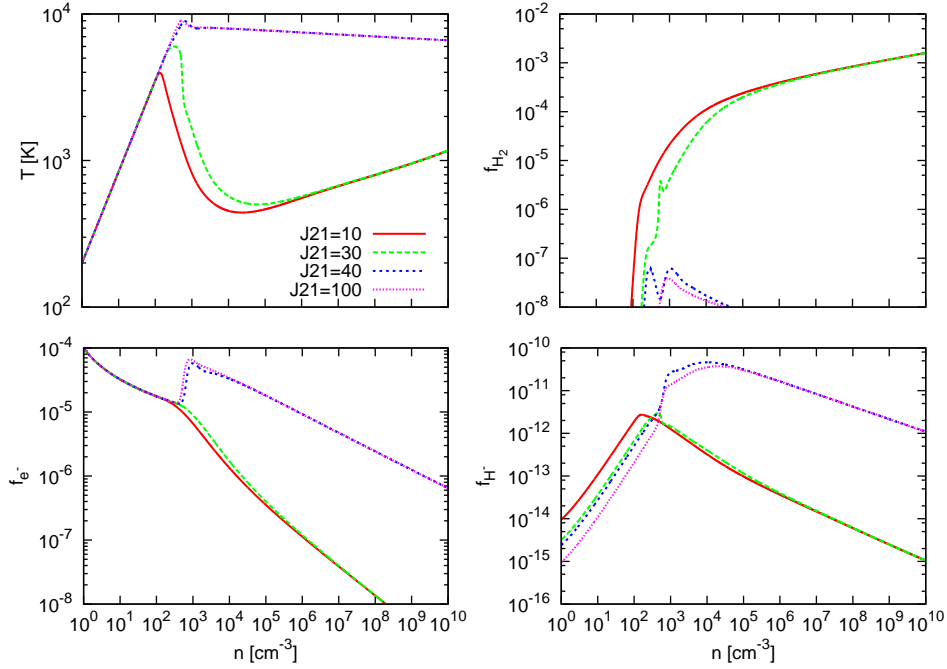}
\end{minipage}
\end{tabular}
\caption{ The abundances of $\rm H_{2}$, $\rm e^{-}$ , $\rm H^{-}$ and temperature are plotted against the number density for various strengths of background UV flux. These results are one-zone calculations are performed using the $\rm H_{2}$ self-shielding function of  Draine \& Bertoldi 1996.}
\label{fig4}
\end{figure*}

\begin{figure*}
\centering
\begin{tabular}{c c}
\begin{minipage}{4cm}
\hspace{-5cm}
\includegraphics[scale=0.45]{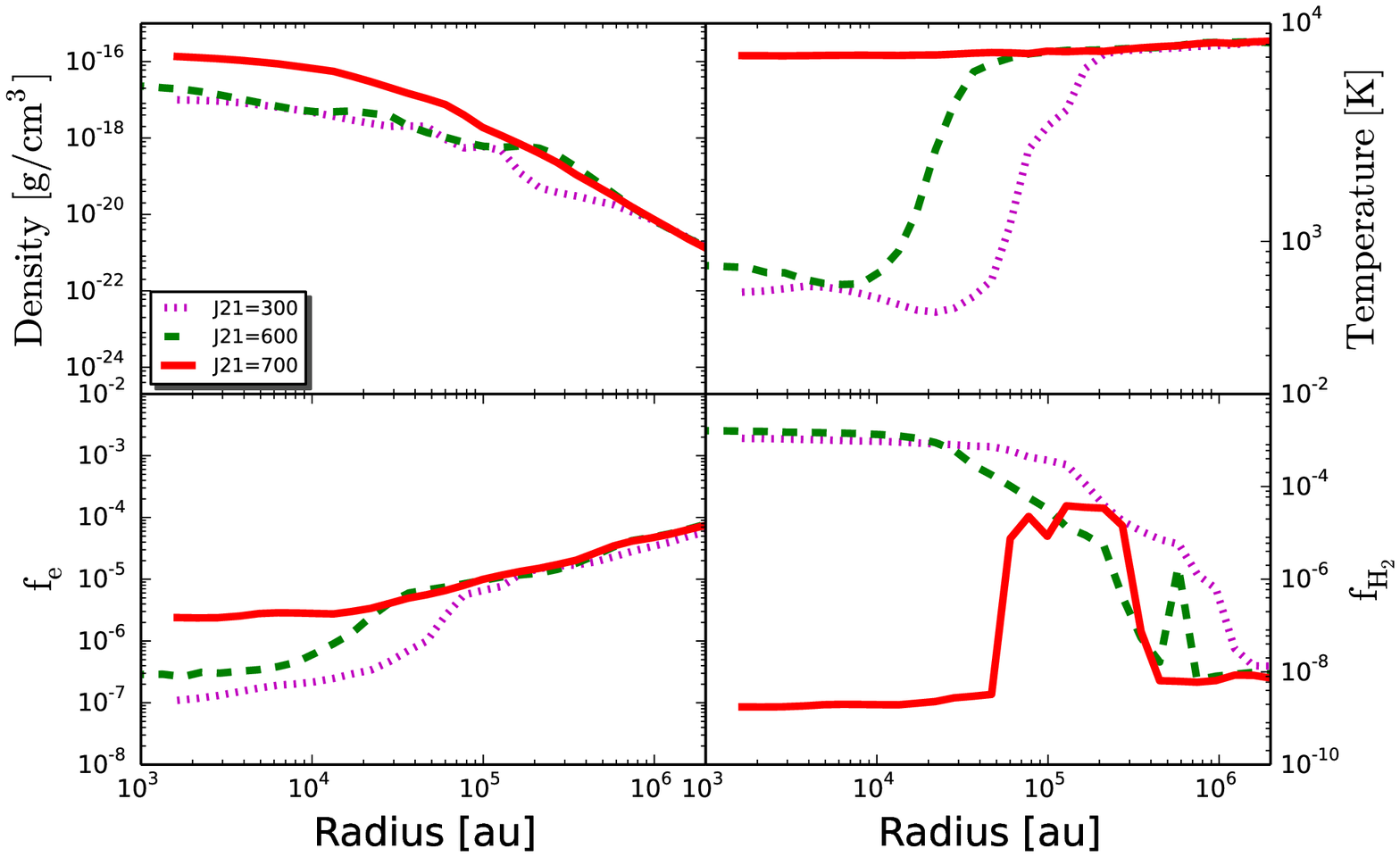}
\end{minipage} &
\begin{minipage}{4cm}
\includegraphics[scale=0.45]{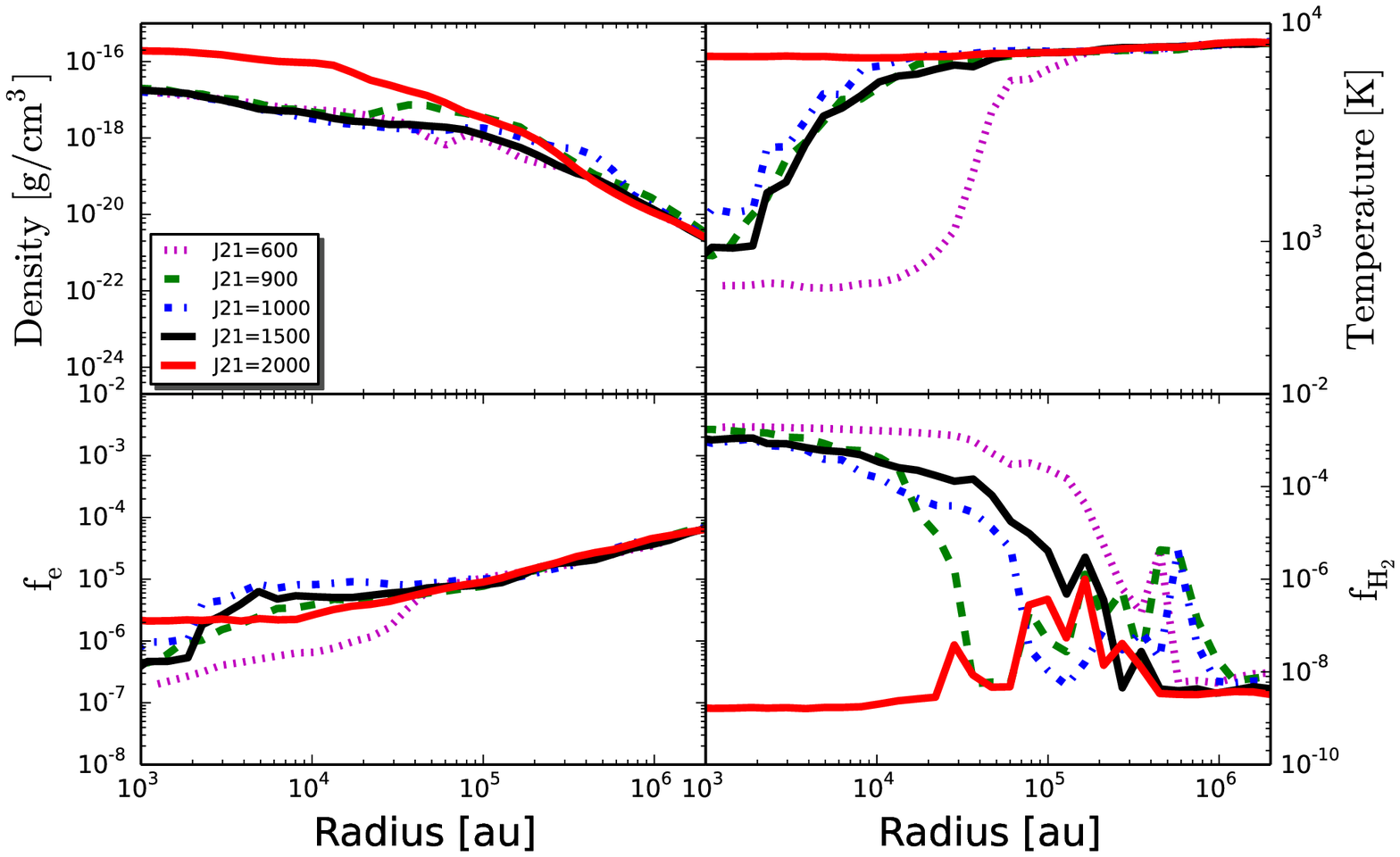}
\end{minipage} \\ \\
\begin{minipage}{4cm}
\hspace{-5cm}
\includegraphics[scale=0.45]{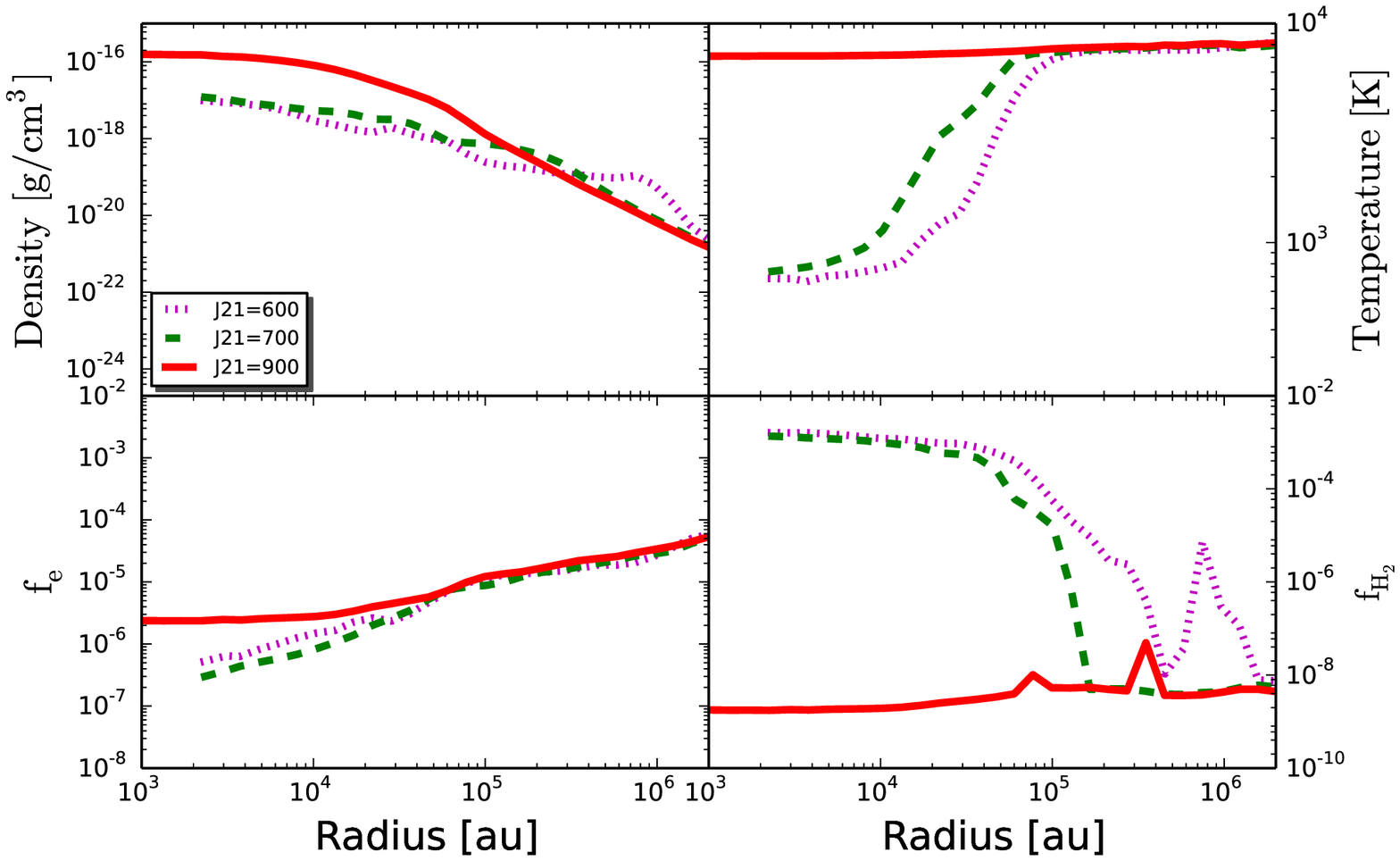}
\end{minipage} &
\begin{minipage}{4cm}
\includegraphics[scale=0.45]{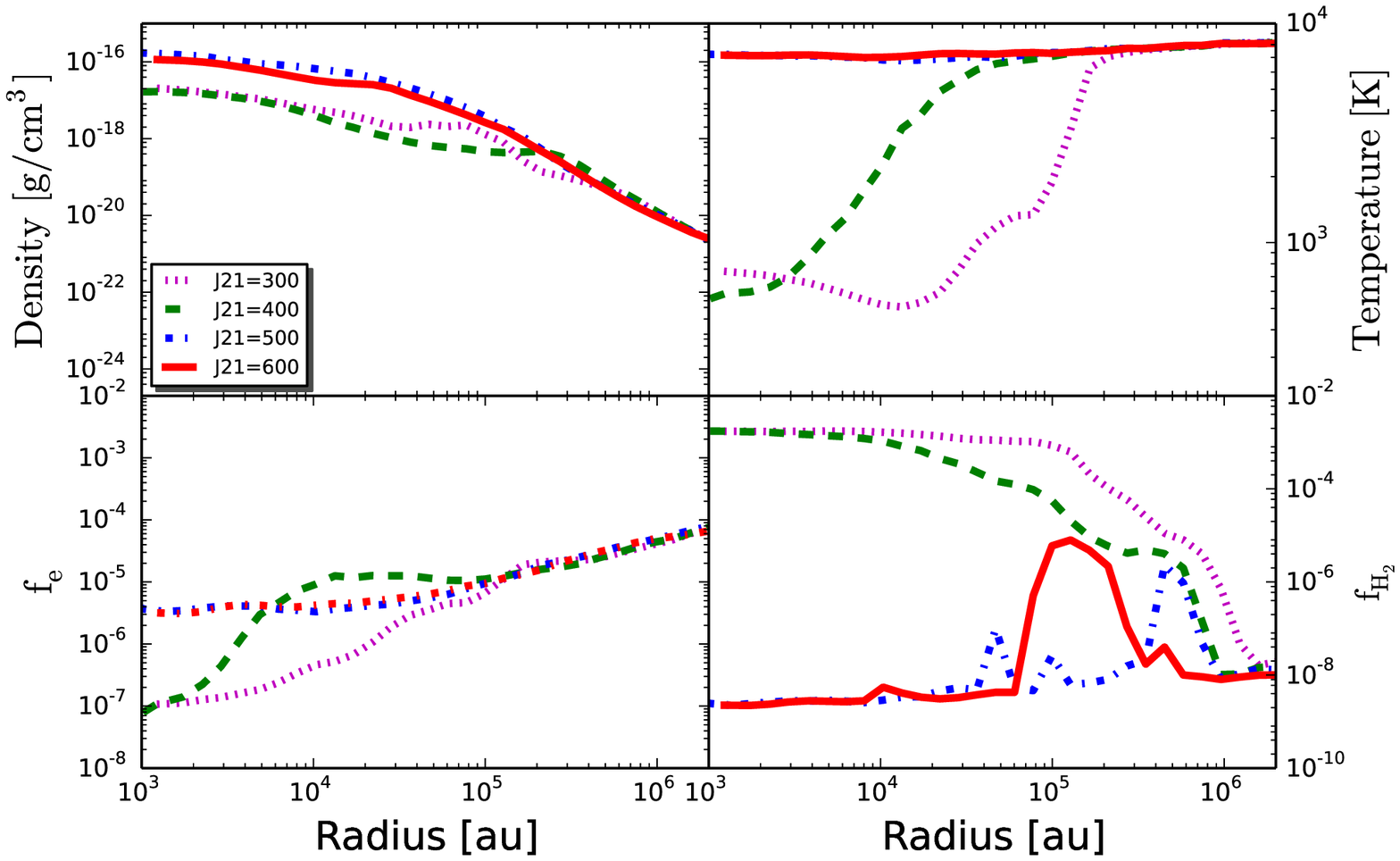}
\end{minipage} \\
\begin{minipage}{4cm}
\includegraphics[scale=0.45]{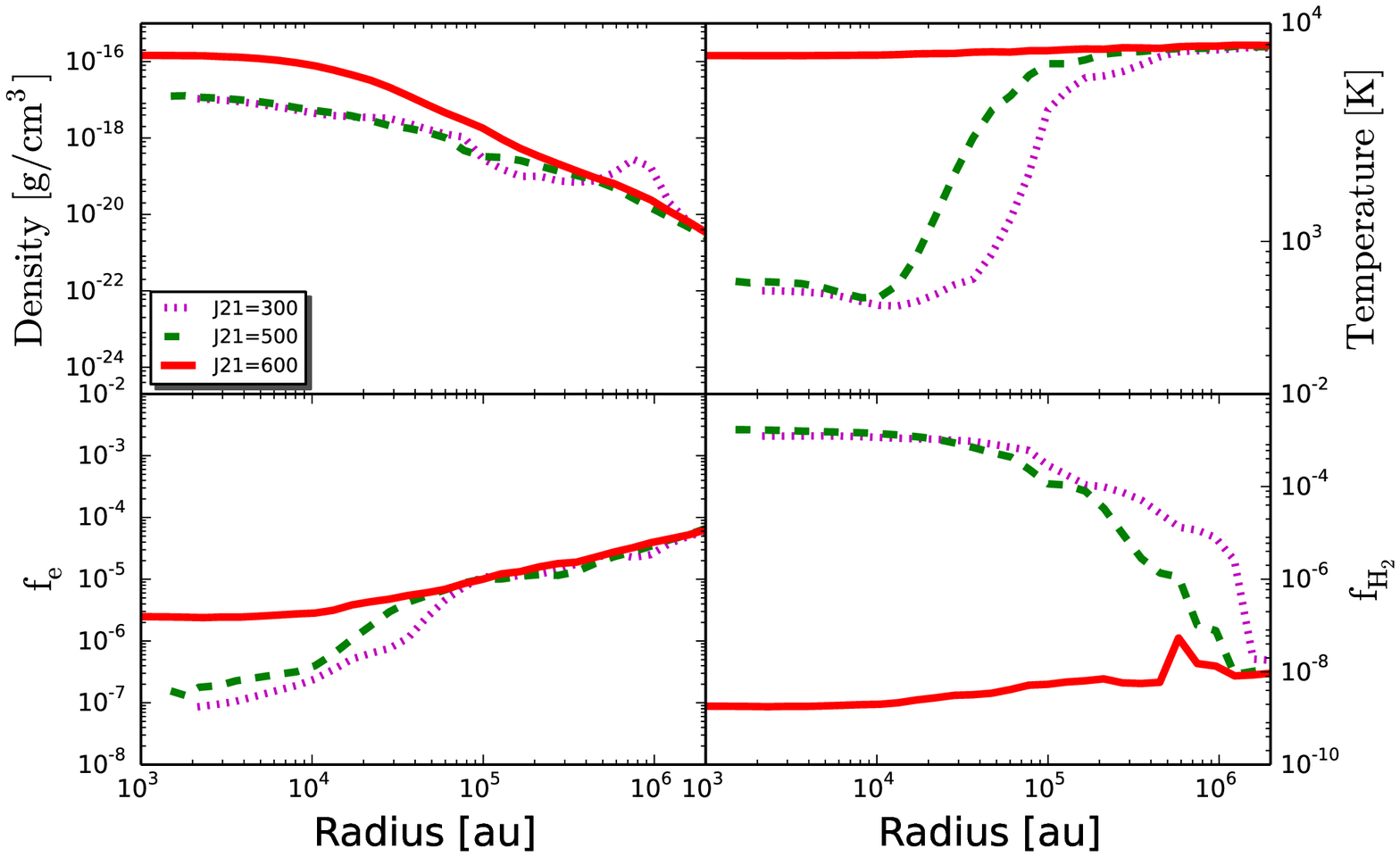}
\end{minipage}
\end{tabular}
\caption{Spherically averaged and radially binned profiles of temperature, density, $\rm H_{2}$ and $\rm e^{-}$ fractions are plotted in this figure. Each panel represent a single halo. Each line style represents the value of $\rm J_{21}$ as mentioned in the legend. Top  panels represent halo A \& B (left to right), middle panels halos C \& D (left to right) and bottom panel halo E. }
\label{fig1}
\end{figure*}

\begin{figure*}
\centering
\begin{tabular}{c c}
\begin{minipage}{4cm}
\hspace{-5cm}
\includegraphics[scale=0.36]{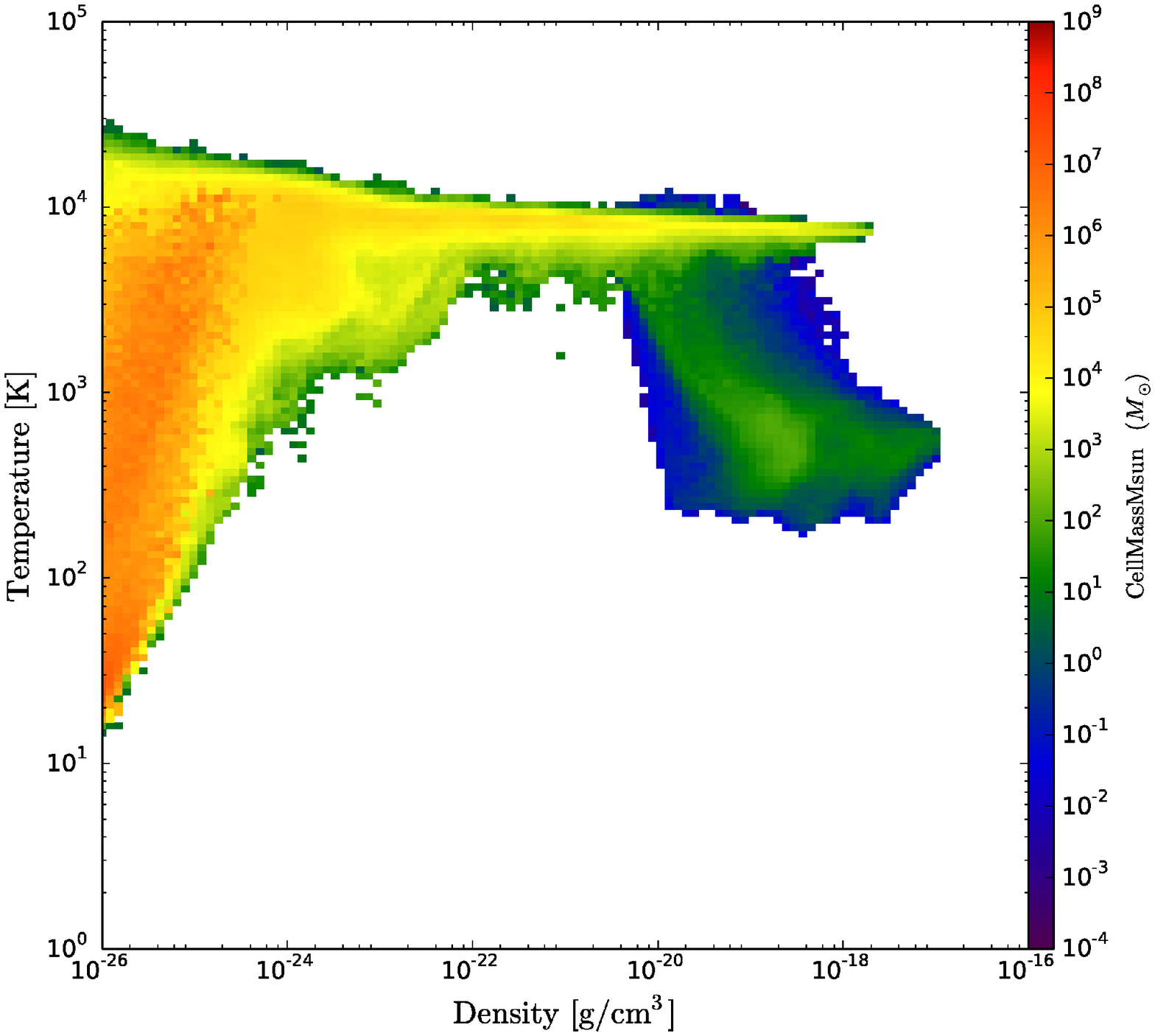}
\end{minipage} &
\begin{minipage}{4cm}
\includegraphics[scale=0.36]{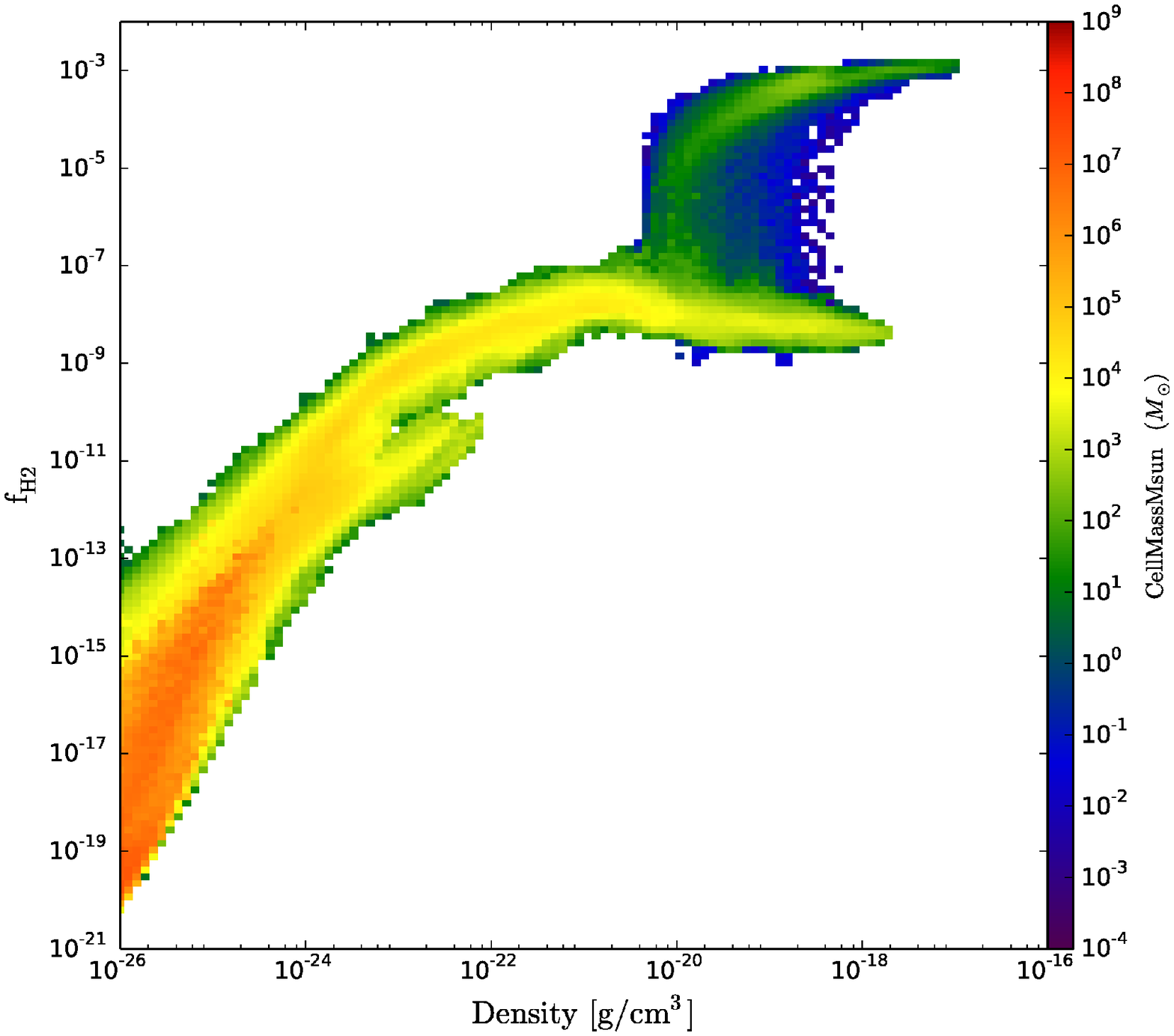}
\end{minipage} \\ \\
\begin{minipage}{4cm}
\hspace{-5cm}
\includegraphics[scale=0.36]{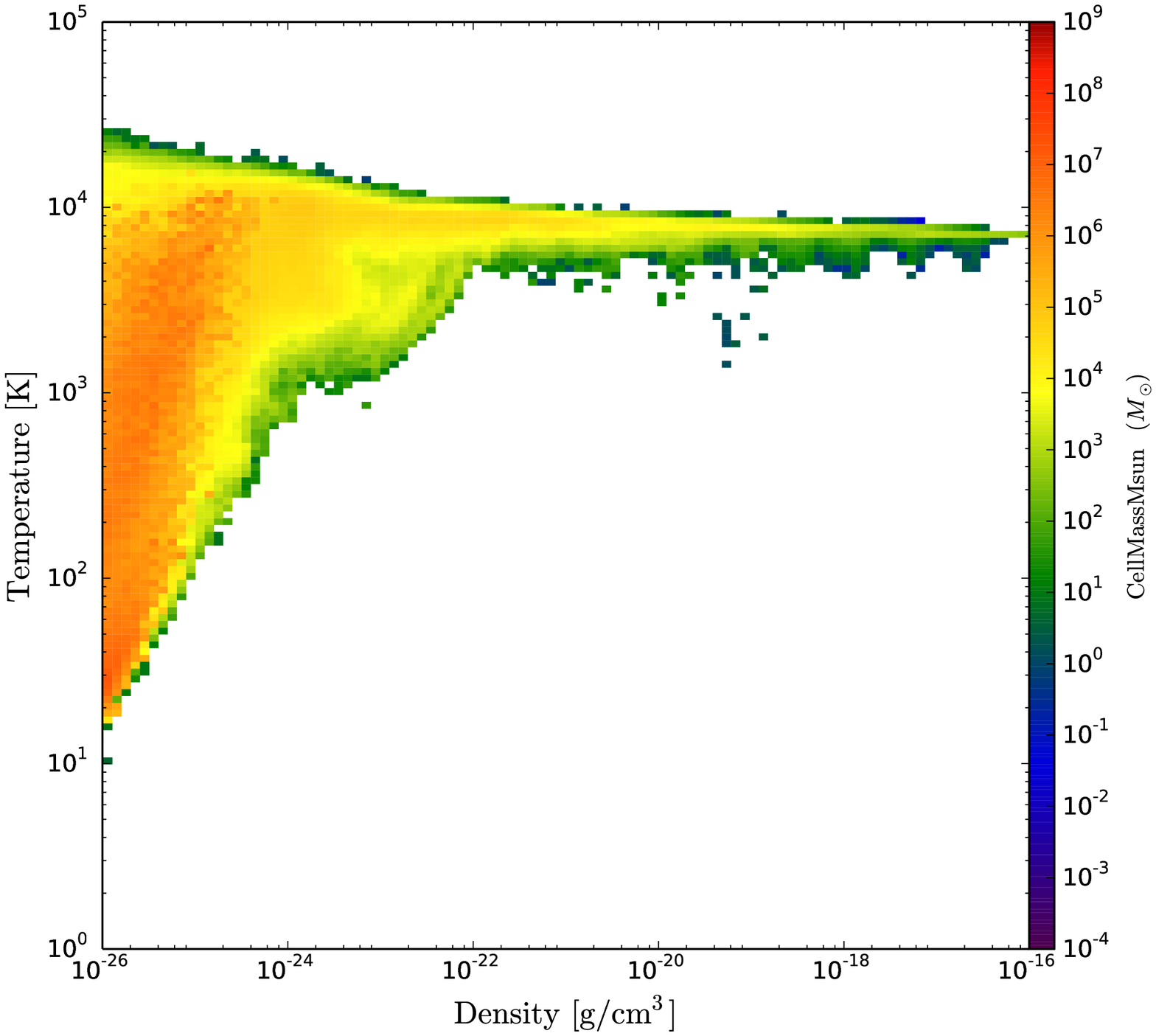}
\end{minipage} &
\begin{minipage}{4cm}
\includegraphics[scale=0.36]{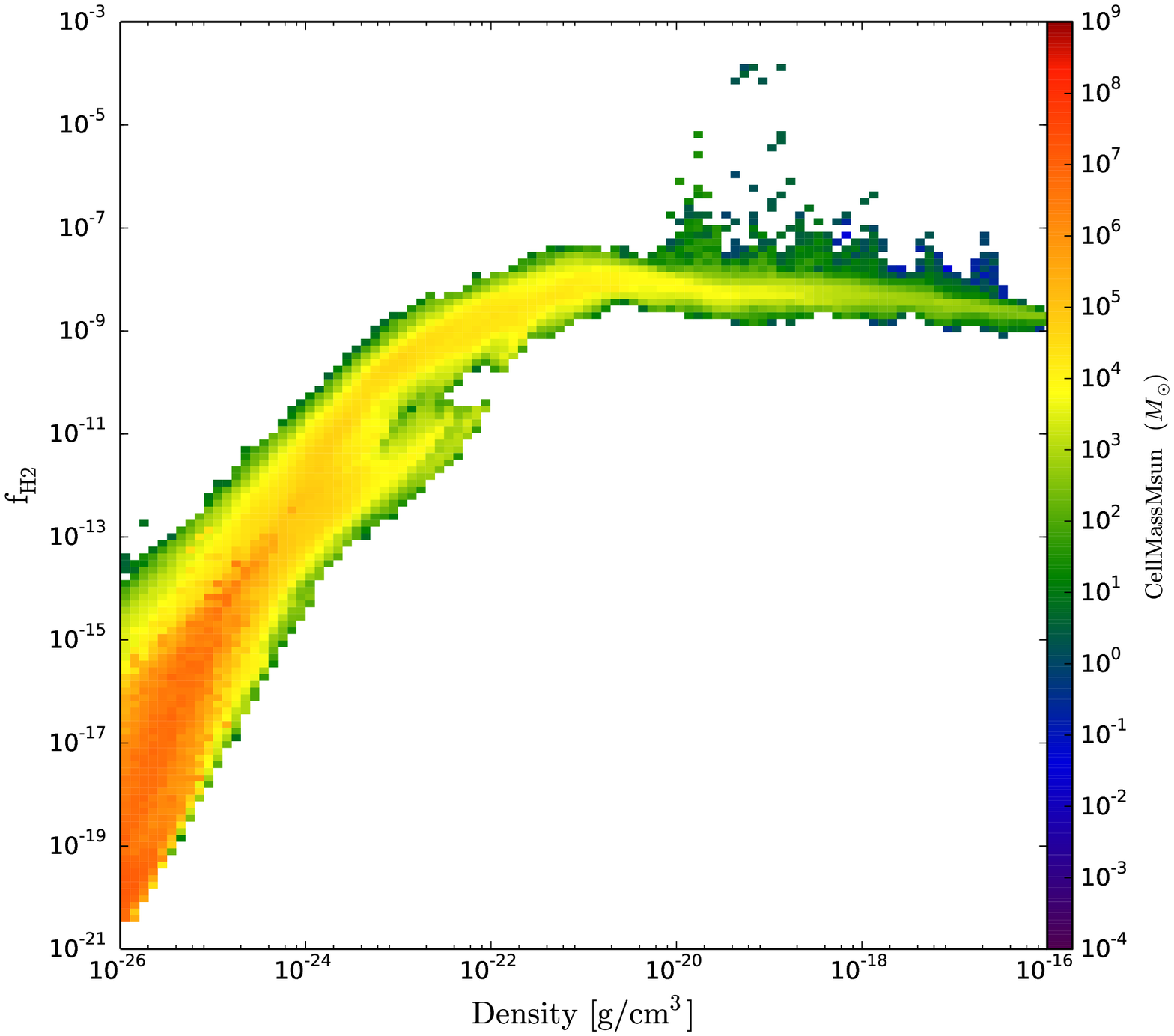}
\end{minipage} \\ \\
\begin{minipage}{4cm}
\hspace{-5cm}
\includegraphics[scale=0.36]{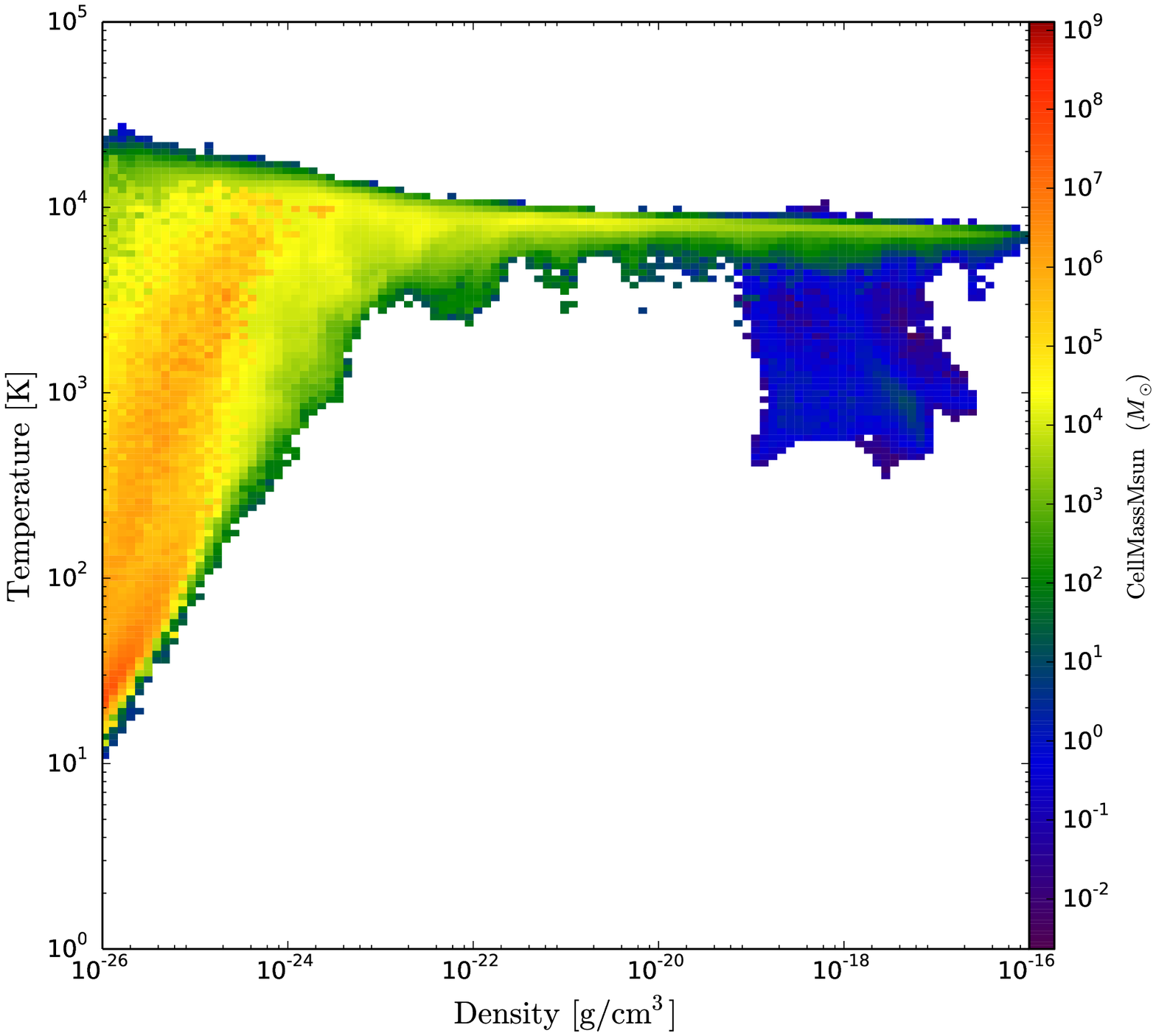}
\end{minipage} &
\begin{minipage}{4cm}
\includegraphics[scale=0.36]{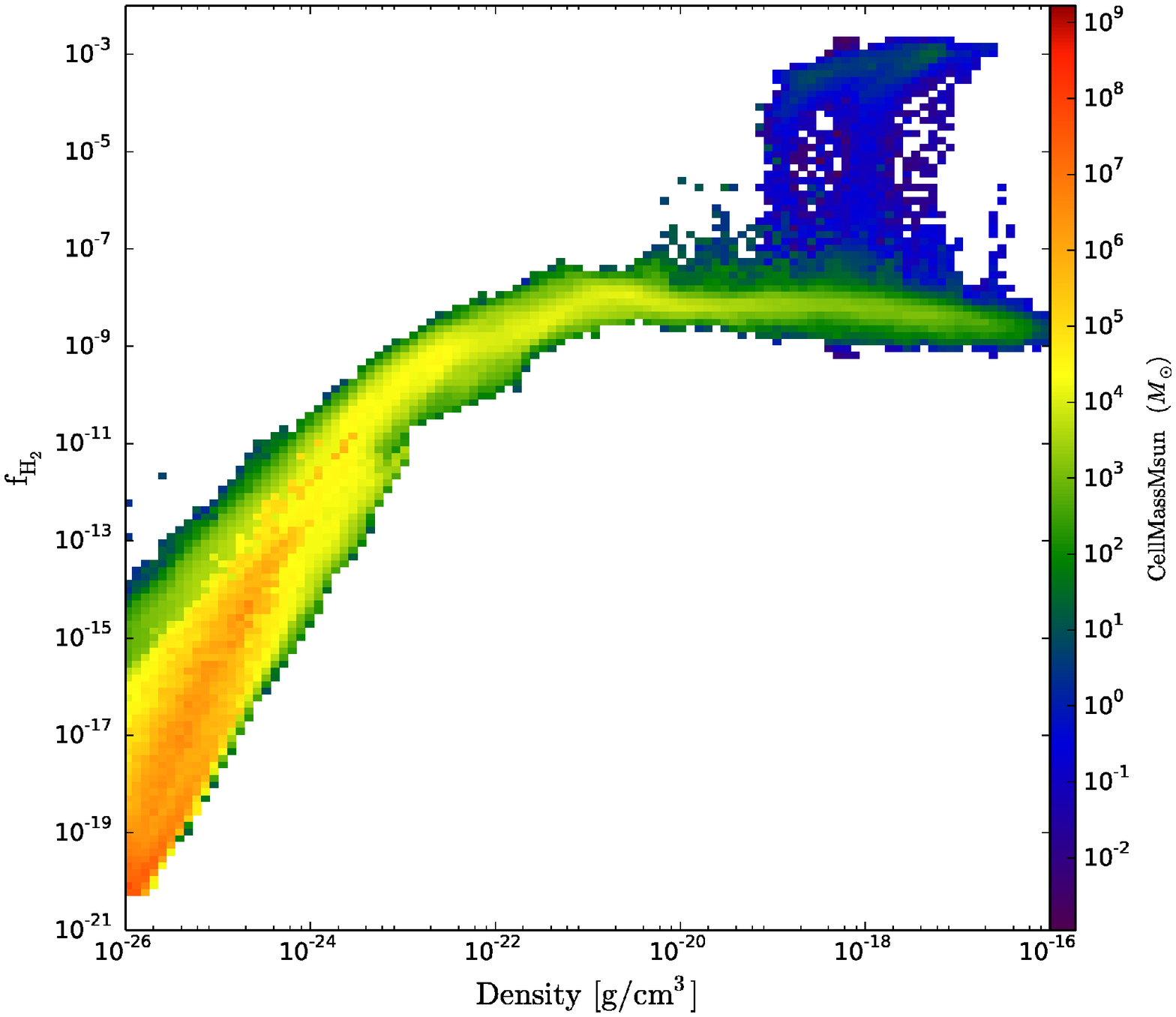}
\end{minipage}
\end{tabular}
\caption{Phase plots of temperature, density and H$_{2}$ fraction for representative cases of halos E and D are shown here. The top panel shows $\rm J_{21}=300$ for halo E, the middle panel $\rm J_{21}=600$ for the halo E, and the bottom panel $\rm J_{21}=600$ for the halo D. The latter case also illustrates the existence of two gas phases at the same density.}
\label{fig11}
\end{figure*}

\begin{table*}
\begin{center}
\caption{Properties of the simulated halos are listed here.}
\begin{tabular}{llll|l|llll|l|llll}
\hline
\multicolumn{4}{c}{halo A} &  \multicolumn{1}{c}{} & \multicolumn{4}{c}{halo B} &  \multicolumn{1}{c}{} & \multicolumn{4}{c}{halo C}\\
\hline
J$_{21}$ & Mass ($\rm M_{\odot} $) &z  &T$_{cent}$(K) & &J$_{21}$ & Mass($\rm M_{\odot} $) &z  &T$_{cent}$(K) & &J$_{21}$ &Mass($\rm M_{\odot}$)	 & z  & T$_{cent}$(K)\\
\cline{1-4} \cline{6-9} \cline{11-14}                                             \\
 		  
 300   &$\rm 1.41 \times 10^{7}$   &14.24    &600   &  &600     & $\rm 2.3 \times 10^{7}$    &12.98   &700  & &600  & $\rm 3.22 \times 10^{7}$  &11.20  &600 \\
 600   & $\rm 1.42 \times 10^{7}$  &14.23    &700   &  &900     & $\rm 2.6 \times 10^{7}$    &12.84   &1000 & &700   & $\rm 3.26 \times 10^{7}$  &11.11  &700 \\
 700   & $\rm 1.43 \times 10^{7}$  &14.22    &7500  &  &1000    & $\rm 2.4 \times 10^{7}$    &12.96   &1200 & &900   & $\rm 3.24 \times 10^{7}$  &11.13  &7500 \\
       &                           &         &      &  &1500    & $\rm 2.5 \times 10^{7}$    &12.97   &1500 & &      &  &   & \\
       &                           &         &      &  &2000    & $\rm 2.4 \times 10^{7}$    &12.93   &7500 & &      &  &   &
 \\
\hline
\multicolumn{4}{c}{halo D} &  \multicolumn{1}{c}{} & \multicolumn{4}{c}{halo E} \\
\cline{1-4} \cline{6-9} 
\\
 300    & $\rm 4.06 \times 10^{7}$   &13.29   &700    &   &300     & $\rm 5.46 \times 10^{7}$  &10.60   &600\\
 400    & $\rm 4.08 \times 10^{7}$  &13.28   &500    &   &500     & $\rm 5.56 \times 10^{7}$  &10.549  &700\\
 500    & $\rm 4.08 \times 10^{7}$  &13.27   &7500   &   &600     & $\rm 5.47 \times 10^{7}$  &10.59   &7500\\
 600    & $\rm 4.1 \times 10^{7}$   &13.24   &7500   \\
  
\hline
\end{tabular}
\label{table0}
\end{center}
\end{table*}

\begin{figure*}
\centering
\begin{tabular}{c}
\begin{minipage}{6cm}
\hspace{-5cm}
\includegraphics[scale=0.8]{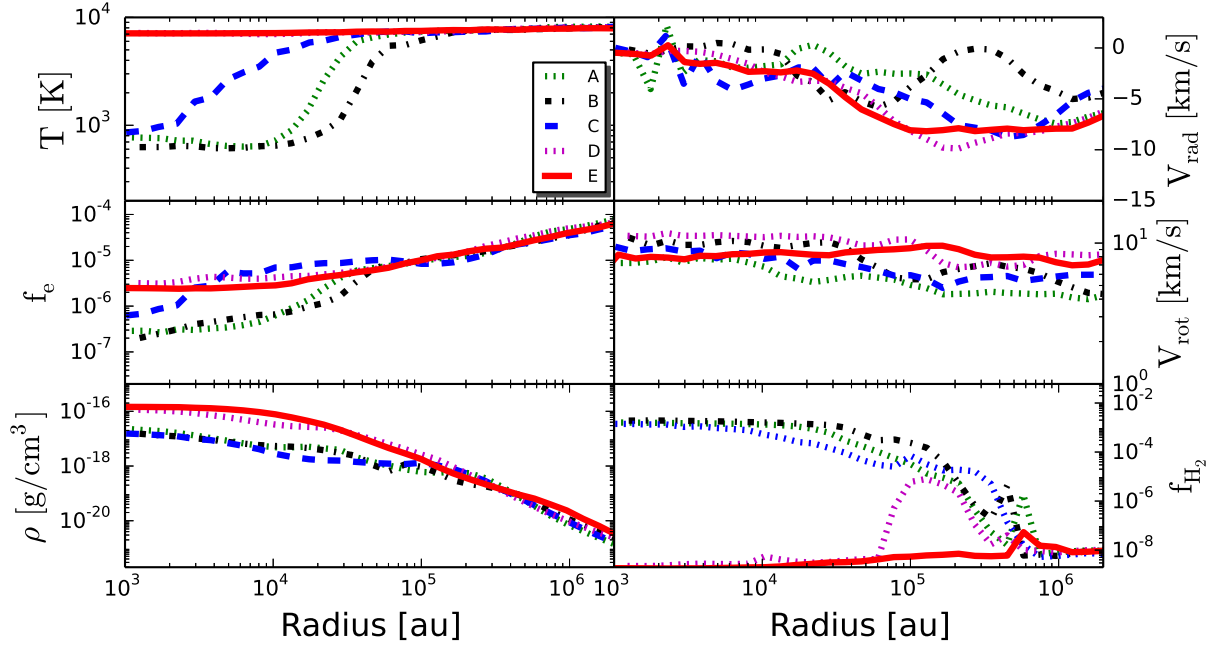}
\end{minipage} 
\end{tabular}
\caption{Spherically averaged and radially binned profiles of temperature, density, $\rm H_{2}$ fraction, $\rm e^{-}$ fraction, radial infall velocity and rotational velocities are plotted for all halos for $\rm J_{21}=600$.}
\label{fig2}
\end{figure*}

\begin{figure*}
 \hspace{-4.0cm}
\centering
\begin{tabular}{c c}
\begin{minipage}{6cm}
\includegraphics[scale=0.4]{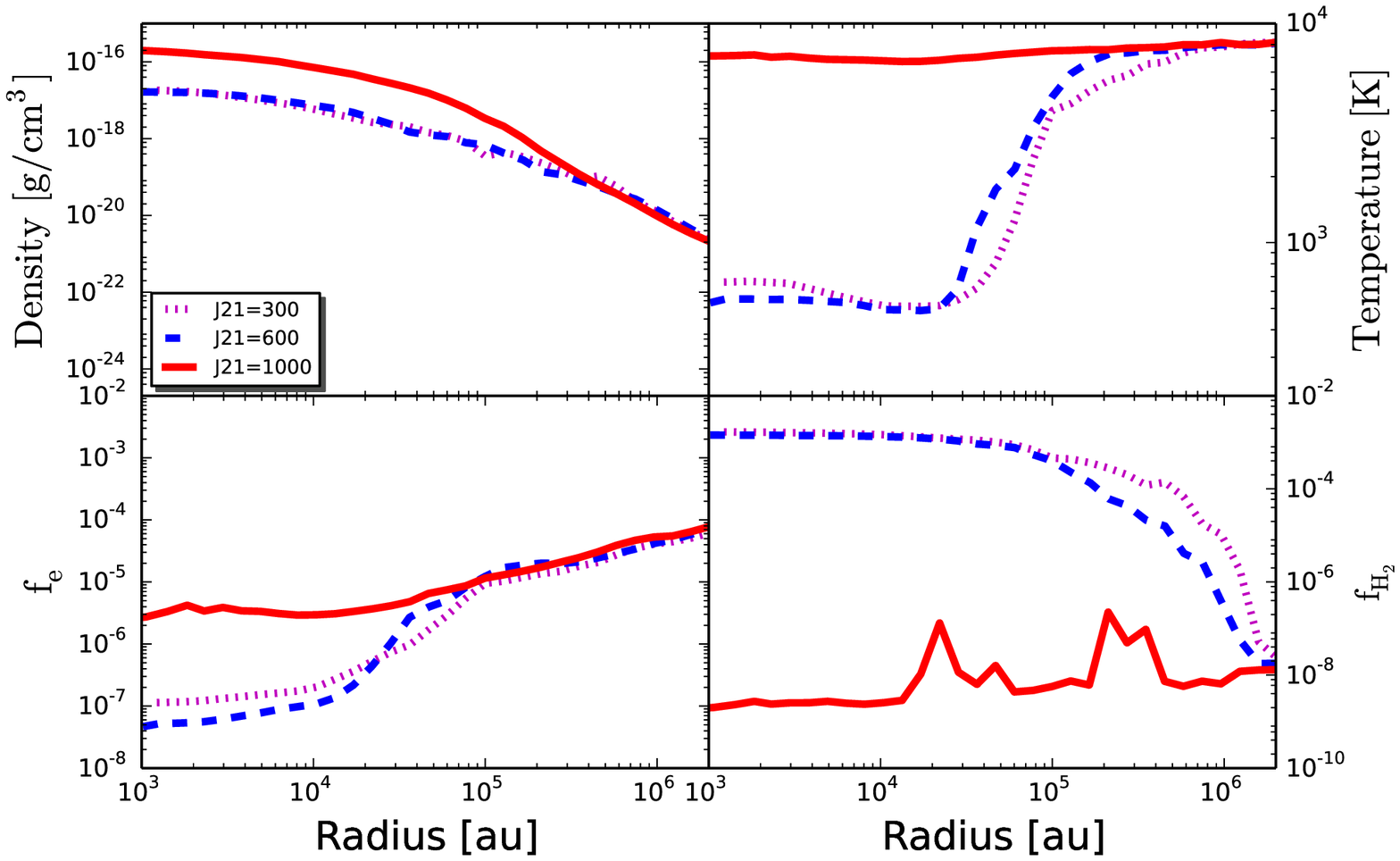}
\end{minipage} &
\hspace{2cm}
\begin{minipage}{6cm}
\includegraphics[scale=0.4]{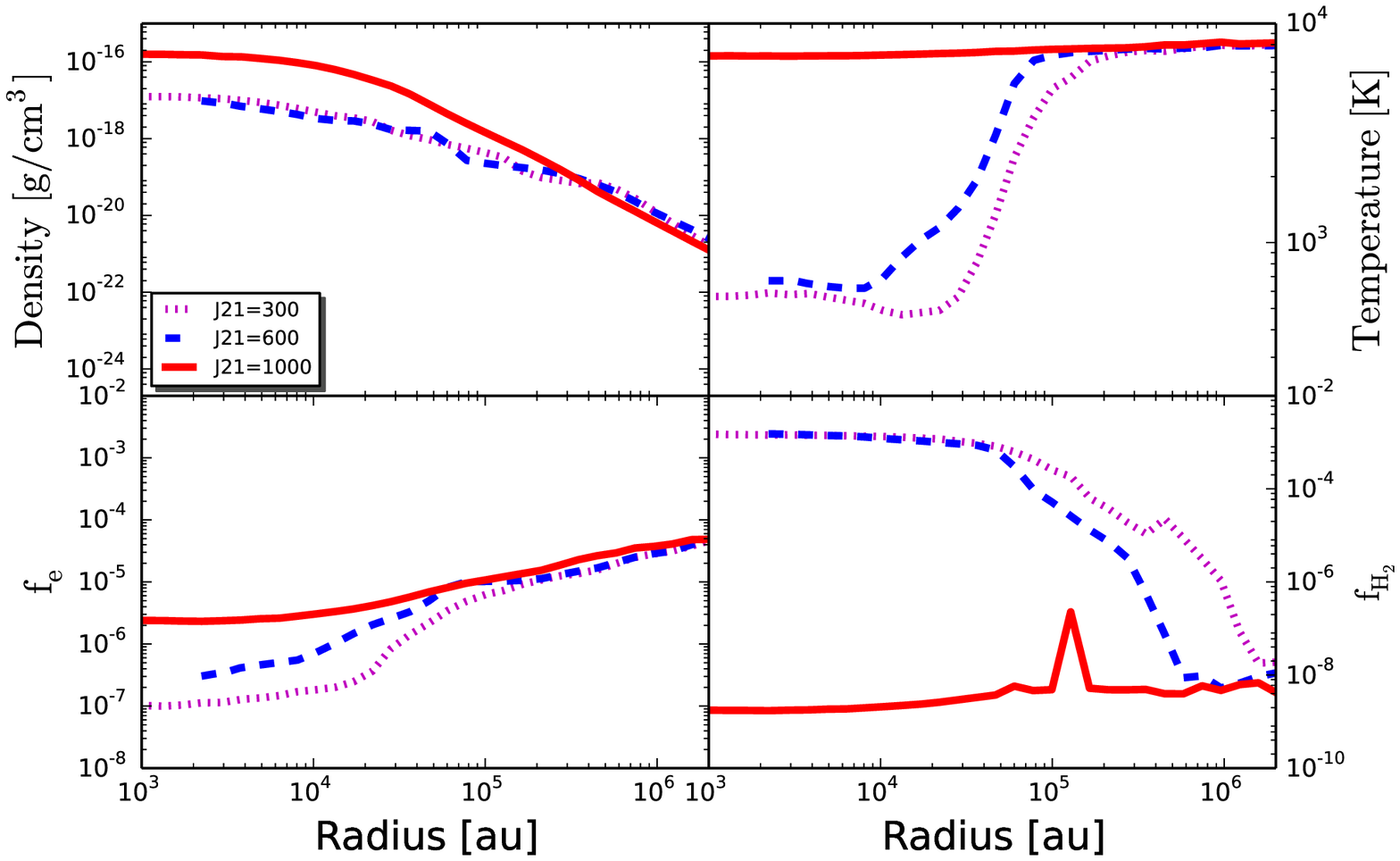}
\end{minipage}
\end{tabular}
\caption{Same as figure \ref{fig1} for halo D \& C but using the $\rm H_{2}$ self-shielding fitting function of Draine \& Bertoldi 1996.}
\label{fig3}
\end{figure*}

\begin{figure*}
 \hspace{-4.0cm}
\centering
\begin{tabular}{c c}
\begin{minipage}{6cm}
\includegraphics[scale=0.4]{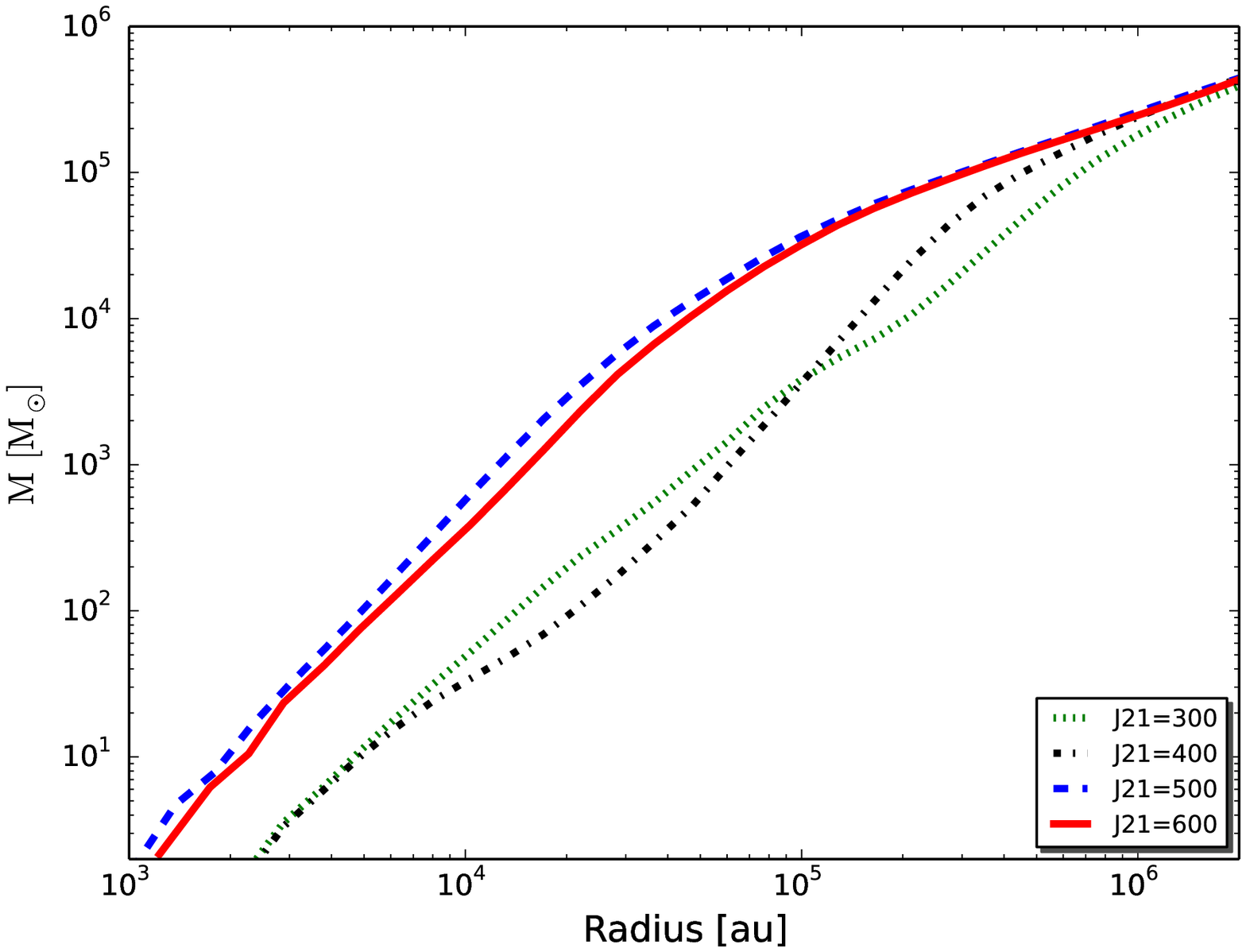}
\end{minipage} &
\hspace{2cm}
\begin{minipage}{6cm}
\includegraphics[scale=0.4]{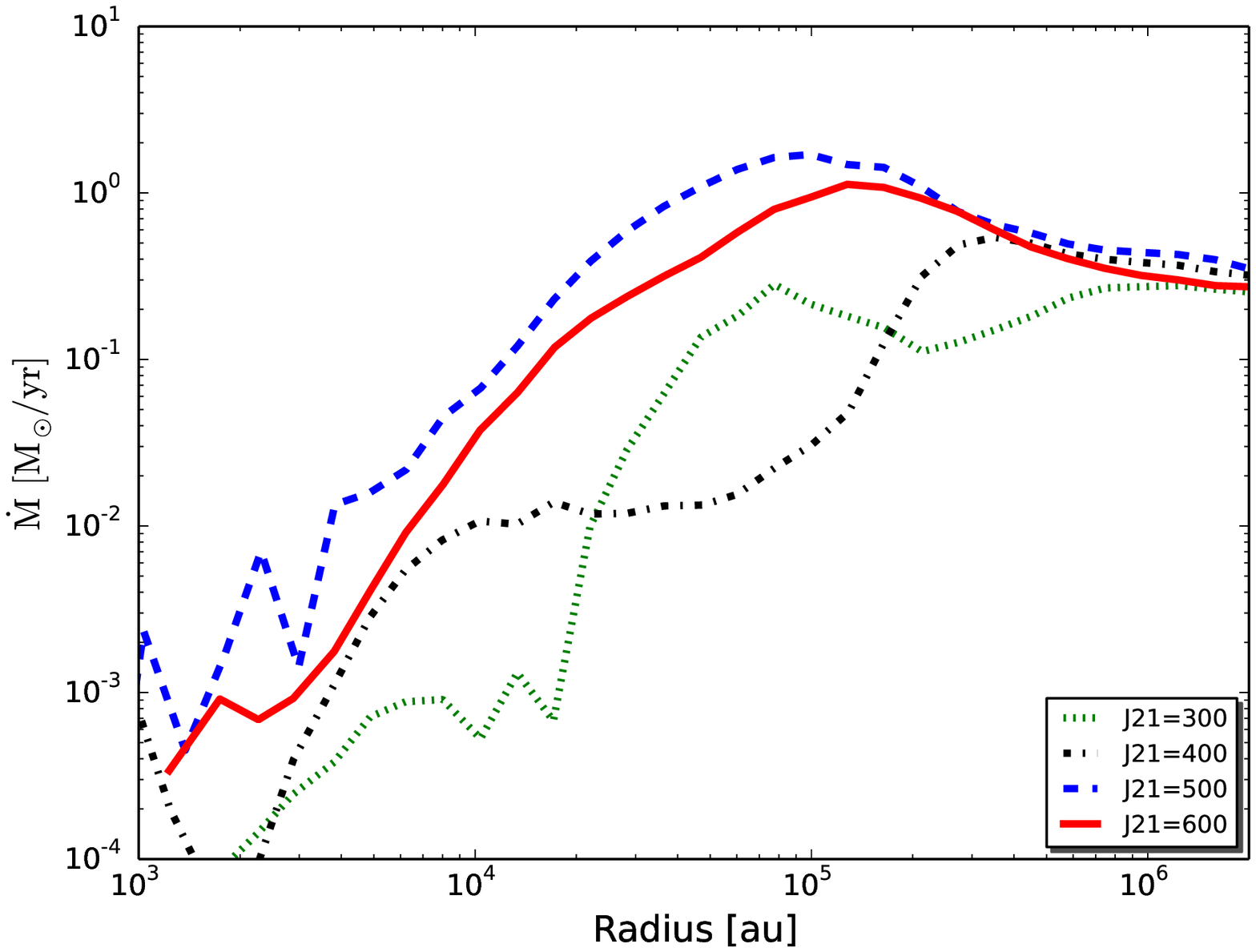}
\end{minipage}
\end{tabular}
\caption{ Enclosed masses and mass accretion rates are shown for a representative case of 'halo D'. The left panel shows enclosed mass profiles while the right panel shows the mass accretion rates.}
\label{fig5}
\end{figure*}

\section{Main Results}

\subsection{Results from one-zone models}
To test our chemical model, we have performed one-zone calculations similar to S10. The thermal evolution and species fractions of  H$_{2}$, H$^{-}$ and e$^{-}$ are shown in figure \ref{fig4}. We start with an initial density of $\rm 1~cm^{-3}$, a gas temperature of 200 K, electron and H$_{2}$ fractions of $\rm 10^{-4}$ and $10^{-10}$ respectively (same as S10). It can be seen from figure \ref{fig4} that for $\rm J_{21}< 30$, H$_{2}$ formation takes place and leads to cooling down to a few hundreds K. The H$^{-}$ fraction initially increases up to densities of 100 cm$^{-3}$, then gets depleted during the formation of H$_{2}$. For stronger fluxes, the formation of H$_{2}$ remains inhibited, cooling due to Lyman alpha photons kicks in and the temperature remains about 8000 K. The thermal evolution and the abundances of the species are in agreement with S10. The value of $\rm J_{21}^{crit}$ found from one-zone calculations is between 30-40 and is consistent with findings of S10. 

However, we note that the value of $\rm J_{21}^{crit}$ differs in 3D simulations. The differences in $\rm J_{21}^{crit}$ between one-zone and three-dimensional simulations  arise from the fact that the one-zone calculation do not capture shocks, collapse dynamics, hydrodynamical effects as well as the effects of the dark matter potential which are crucial in determining the $\rm J_{21}^{crit}$.

\subsection{Results from 3D simulations}
To determine $\rm J_{21}^{crit}$, we have performed more than 20 cosmological simulations for five distinct halos with masses above 10$^{7}$~M$_{\odot}$ including all the necessary processes for the formation and dissociation of molecular hydrogen by self-consistently solving the rate equations along with the hydrodynamics. In figure \ref{fig1}, we present the spherically averaged profiles of temperature, density, $\rm H_{2}$ and $\rm e^{-}$ abundances for all simulated halos. It is found that in all halos the formation of $\rm H_{2}$ takes place for $\rm J_{21} < 300$. At radii $\rm > 2 \times 10^5$ au, the temperature in all cases for the various strengths of $\rm J_{21}$ is about 7500 K but the fraction of the molecular hydrogen remains higher for the lower values of $\rm J_{21}$. This is because the $\rm H^{-}$ photo-detachment rate scales linearly with the strength of $\rm J_{21}$, i.e. the stronger the flux the lower the fraction of molecular hydrogen. For fluxes below the critical values, the fraction of $\rm H_{2}$ gets further boosted during the collapse and reaches $\rm \sim10^{-3}$. It cools the gas down to temperatures of about thousand K. Similarly, the electron fraction declines towards the center. It is about $\rm 10^{-7}$ in the center and  $\rm 10^{-4}$ at larger radii. The impact of H$_{2}$ cooling is also visible in the density profile which significantly deviates from the isothermal profile. Small bumps in the density profile are typical signatures of fragmentation which is expected in these cases. This trend is observed in all the halos for $\rm J_{21} = 300$.

For some halos the fraction of H$_{2}$ sufficient to induce cooling still occurs at later times for $\rm J_{21} > 300$, at densities of about $\rm 10^5~cm^{-3}$. This lowers the halo central temperatures down to about a thousand K. For $\rm J_{21}$ above the critical threshold the fraction of $\rm H_{2}$ remains quite low, i.e. $\rm 10^{-8}$. Such a low fraction is not sufficient to trigger H$_{2}$ cooling and then the halos collapse isothermally. We note that value of the $\rm J_{21}^{crit}$ varies from 400-700 for four halos as listed in table \ref{table1}. A notable variation in the critical value is observed for halo B where $\rm J_{21}^{crit}=1500$. Such variations from halo to halo are not surprising as halos have different density distributions after virialization, different spins and accretion shocks at different densities. As shown in figure 6 \& 7 of S10, collisional dissociation is the main destruction channel for $\rm H_{2}$ above densities of $\rm 10^3~cm^{-3}$. The rate of collisional dissociation exponentially depends on the temperature \citep{1996ApJ...461..265M}, and small local variations in the temperature can change $\rm J_{21}^{crit}$ as verified from our one-zone model by artificially changing the temperature by few hundreds K. Additional fluctuations in the electron number density may further influence it and there is a weak dependence of the collisional dissociation on density as well. As these quantities change simultaneously when different halos are considered, the individual dependencies cannot be explored in isolation, but the functional form of collisional dissociation rate suggests that temperature dependence provides the dominant effect (also see discussion in S10).

In figure \ref{fig11} we show the phase plots of temperature and $\rm H_{2}$ fraction against density. Initially, the gas is heated up to the virial temperature of the halo, i.e. above $\rm 10^{4}~K$, and then cools down to about 8000 K by Lyman alpha radiation. At densities of about $\rm 10^{4}~cm^{3}$, cooling due to H$_{2}$ becomes effective for the weaker $\rm J_{21}$ case while for the stronger flux case it remains inhibited. It is found that temporarily different gas phases may coexist at the same density. The latter reflects local variations in gas density, temperature and self-shielding. It may also be noted that in the latter case, the amount of gas in the cold phase is negligible compared to the top panel which has a significant amount of cold gas and remains in the cold phase. In these simulations, we find that if the gas at densities $\rm \geq 10^{3}~cm^{3}$ has a temperature lower than 1000 K (as shown in top panel of \ref{fig11}) then cooling due to $\rm H_{2}$ becomes important and halos remains in cold phase. On the other hand if gas at densities $\rm \geq 10^{3}~cm^{3}$ has a temperature higher than 2000 K shown in the bottom panel of figure \ref{fig11} then the halo remains in the hot phase and collapses isothermally. We have also checked that the behavior remains the same if the simulations are evolved to higher densities.


To further understand the origin of $\rm J_{21}^{crit}$, we have plotted the radially averaged density, temperature, abundances of $\rm H_{2}$ and $\rm e^{-}$, radial infall velocities as well as rotational velocities of all halos for $\rm J_{21}=600$ in figure \ref{fig2}. For the given strength of $\rm J_{21}$ two halos (D \& E) are already in an isothermal state with central temperature around 7000 K while the other three have a sufficient $\rm H_{2}$ fraction to reduce their central temperatures down to about 1000 K. It can be noted that particularly halos D \& E have larger infall velocities already at radii $\rm > 10^{5}~au$ compared to the halos with lower central temperatures. Particularly, the halo B with the highest $\rm J_{21}^{crit}$ in our sample has the lowest infall velocity. The differences in the infall velocities arise due to the ambient sound speed of gas cloud.
We further note that the two most massive halos have a lower value of $\rm J_{21}^{crit}$ than the lower-mass halos. S10 have considered even more massive halos at lower redshift, finding a further decrease in $\rm J_{21}^{crit}$. The latter suggests a potential dependence on the mass of the halo with some fluctuations.

\begin{table}
\begin{center}
\caption{Properties of the simulated halos for $\rm J_{21}^{crit}$ are listed here.}
\begin{tabular}{ccccccc}
\hline
\hline

Model	& Mass	 & Redshift  & $\rm J_{21}^{crit}$  & spin parameter\\

No & $\rm M_{\odot} $   &z  &  in units of J$_{21}$  &$\lambda$ \\
\hline                                                          \\

A     & $\rm 1.42 \times 10^{7}$  &14.23  &600  &0.025\\
B     & $\rm 2.4 \times 10^{7}$   &12.95  &1500 &0.009\\
C     & $\rm 3.25 \times 10^{7}$  &11.11  &700  &0.03\\
D     & $\rm 4.06 \times 10^{7}$  &13.29  &400  &0.02\\
E     & $\rm 5.6 \times 10^{7}$   &10.55  &500  &0.034\\

\hline
\end{tabular}
\label{table1}
\end{center}
\end{table}

\subsection{Comparison with previous studies}
For a comparison of our results with S10, we have performed a couple of simulations where we employed exactly the same chemical and thermal processes as described by S10 without any modifications and using the $\rm H_{2}$ self-shielding fitting function of \cite{1996ApJ...468..269D}. These simulations were performed for halos C \& D and are shown in figure \ref{fig3}. 
It suggests that the value of $\rm J_{21}^{crit}$ differs by a factor of a few from their estimates as they suggest $\rm J_{21}^{crit}$ varies from 30-300. Our study brackets the value of $\rm J_{21}^{crit}$ within a factor of two. The self-shielding fitting function of \cite{1996ApJ...468..269D} is known to overestimate the $\rm H_{2}$ shielding. Therefore, a more accurate determination of $\rm J_{21}^{crit}$ is not necessary. In addition, we found significant variations from halo to halo as well. Given the strong dependence of $\rm J_{21}^{crit}$ on the variations in the local gas temperatures, halo merger histories, the occurrence of shocks at various densities as well as the density and temperature dependence of the $\rm H_{2}$ collisional dissociation rates, the results may still be consistent with S10, even though we typically obtain higher values for the $\rm J_{21}^{crit}$. Variations in the temperature and the occurrence of the shocks is evident from figure \ref{fig11}. As mentioned in previous section, we further verified it by artificially changing the temperature in our one-zone model. This was already confirmed by S10 as well (see paragraph 2, section 3.5 of S10). It may also be noted that the halos in our sample are assembled at higher redshifts and have lower masses compared to the S10, and therefore presumably different structures as well as formation histories.

We further noted that the value of $\rm J_{21}^{crit}$ very weakly depends on the choice of self-shielding function. It is due to the fact that for a stellar temperature of T$_{*}=10^{4}$~K the main dissociation channel is H$^{-}$ photo-detachment not the direct dissociation of H$_{2}$. The value of $\rm J_{21}^{crit}$ changes within a factor of two at-most by employing the self-shielding fitting function of \cite{1996ApJ...468..269D} which overestimates the shielding effect. S10 also found similar results (private communication with Zoltan Haiman).


\subsection{Implications for the formation of DCBHs}

The key requirements for the formation of DCBHs are that the gas in halos with $\rm T_{vir}> 10^{4}~K$ must be of primordial composition and the formation of molecular hydrogen remains suppressed to avoid fragmentation. To keep the gas free from $\rm H_2$ requires the presence of a strong UV flux above the critical value. Such values of $\rm J_{21}^{crit}$ can be achieved in the surrounding of star-burst galaxies or even in the halos which are satellites of such galaxies \citep{2008MNRAS.391.1961D,2012MNRAS.425.2854A,2014arXiv1403.5267A}. We have computed the enclosed mass and mass accretion rates for halo ``D'' as a representative case which is shown in figure \ref{fig5}. It is found that for an isothermal collapse, the gas in the halo remains hotter, leads to higher accretion rates and consequently the enclosed mass within $\rm 10^5~au$ is about two orders of magnitude higher compared to the case with the weaker UV flux. The mass accretion rate peaks around 1~M$_{\odot}$/yr for isothermal cases and overall is about an order of magnitude higher compared to the $\rm H_{2}$ cooling cases. One may expect similar differences in the mass of resulting objects. These enhanced accretion rates help in rapidly building up the DCBHs.  

Our estimates of $\rm J_{21}^{crit}$ are highly relevant for computing the number density of DCBHs and their comparison with observations. Some recent studies predict the number density of DCBHs of few per comoving $\rm Mpc^{-3}$ compared to the observed SMBHs density of few per comoving $\rm Gpc^{-3}$\citep{2008MNRAS.391.1961D,2012MNRAS.425.2854A}. Our results suggest that for the most of halos simulated in these studies $\rm J_{21}^{crit} \geq 400$. The choice of such critical values changes the number density of DCBHs and also helps these models to have better agreement with observations. Based on the prescription for estimating the fraction of halos exposed to super-critical UV flux by \cite{2008MNRAS.391.1961D} and \cite{2012MNRAS.425.2854A}, we estimate that fraction of  the halos hosting DCBHs may be reduced by about three orders of magnitude due to the increase in $\rm J_{21}^{crit}$ by an order magnitude, i.e. adapting $\rm J_{21}^{crit}=400$. The expected number density of BHs is $\rm 10^5$ per comoving $\rm Gpc^{-3}$ at $z=6$. The latter is still factor of 500 or so higher than observed number density. These estimates do not include the metal enrichment by supernova driven winds (see \cite{2008MNRAS.391.1961D} and \cite{2012MNRAS.425.2854A}) which will further reduce the number density of DCBHs. We also report here a potential mass dependence of $\rm J_{21}^{crit}$, which tends to increase with decreasing halo mass. In order to give black holes more time to accrete, the initial collapse should take place at higher redshift, at a virial temperature of $\rm 10^4$ K. The latter implies a decreasing mass for higher-redshift dark matter halos, corresponding to a higher value of  $\rm J_{21}^{crit}$.

\section{Discussion}

One of the main obstacles for the formation of direct collapse black holes is to avoid fragmentation in massive primordial halos which are the potential birthplaces of seed black holes. This may only be possible in the absence of molecular hydrogen which may induce fragmentation and trigger star formation. The ubiquity of background UV flux can photo-dissociate $\rm H_{2}$ molecules and may overcome this obstruction. The prime objective of this work is to determine the critical value of the background UV flux required to suppress the molecular hydrogen formation in atomic cooling halos. As photo-dissociation of $\rm H_{2}$ depends on the type of stellar spectrum, we have considered here UV photons below 13.6 eV emitted by PopII stars. We have conducted three-dimensional cosmological simulations for five distinct halos by including all relevant processes for the formation and dissociation of $\rm H_{2}$. The halos studied here have typical masses of a few times 10$^7$~M$_{\odot}$ and were illuminated by various strengths of background UV flux. We here employed the $\rm H_{2}$ self-shielding fitting function provided by WG11.

Our findings show that the value of $\rm J_{21}^{crit}$ strongly depends on the properties of the halo and may vary from halo to halo. For the halos studied here, we found that the value of $\rm J_{21}^{crit}$ varies from 400-700 with the exception of one halo where it is about 1500. It is also found that $\rm J_{21}^{crit}$ may depend on the mass of halo,  as the two most massive halos have reduced values of $\rm J_{21}^{crit}$. This trend is consistent with results by S10, where more massive halos at have even lower values of $\rm J_{21}^{crit}$. We note that our one-zone calculations are in agreement with S10. The highly non-linear collapse dynamics leads to the occurrence of shocks with Mach numbers of about 3 at various densities which changes the local gas temperature and consequently $\rm J_{21}^{crit}$ differs from halo to halo due to the strong dependence of H$_{2}$ collisional rate on temperature and density \citep{1996ApJ...461..265M}. To build up the supermassive black holes at z $~$6-7, one should preferentially consider halos that collapse early, implying a lower mass and potentially higher $\rm J_{21}^{crit}$ at the same virial temperature. However, it will be desirable in the future to verify it for a larger sample of halos with broader mass range.
Our estimates for $\rm J_{21}^{crit}$ are quite robust as we consider a larger sample of halos and higher Jeans resolution leading to better resolved shocks and employed the high order chemical solver DLSODES. We also find that the value of  $\rm J_{21}^{crit}$ weakly depends on the choice of $\rm H_{2}$ self-shielding. Although, the fitting formula of \cite{1996ApJ...468..269D} overestimates $\rm H_{2}$ self-shielding compared to WG11,  its impact is very low for the adapted stellar spectra.
 
The value of $\rm J_{21}^{crit}$ is about an order of magnitude higher in 3D calculations compared to the one-zone results. This is because of the inability of one-zone calculations to model shocks and hydrodynamical effects.  Similar results have been found in the study of S10. We also included the effect of dissociative tunneling and found from the one-zone test that it decreases $\rm J_{21}^{crit}$ by a factor of three. The estimates of $\rm J_{21}^{crit}$ determined in this work have important implications for the formation of DCBHs. Our results suggest that the value of $\rm J_{21}^{crit}>400$ should be employed in computing the number density of DCBHs. The value of $\rm J_{21}^{crit}$ used in previous studies \citep{2012MNRAS.425.2854A,2014arXiv1403.5267A} seems rather low (i.e. 30) and may be one of the reasons for the high abundance of DCBHs predicted from semi-analytical calculations. From our results, the expected BH number density is ~$\rm 10^5$ per comoving $\rm Gpc^{-3}$ at z= 6 for $\rm J_{21}^{crit}=400$. This estimate is obtained by rescaling the values of \citep{2012MNRAS.425.2854A} for a higher $\rm J_{21}^{crit}$. These estimates do not take into account the metal enrichment in the intergalactic medium by supernova driven winds which may further reduce the number density of DCBHs. 

In our previous studies \citep{2013MNRAS.433.1607L,2013MNRAS.436.2989L,2014MNRAS.440.2969L}, we have shown that the presence of strong UV flux leads to an isothermal collapse where conditions are fertile for the formation of DCBHs. In fact, under these conditions large accretion rates of $>1$~M$_{\odot}$/yr are observed which result in the formation of supermassive stars of $\rm 10^{5}~M_{\odot}$, the potential progenitors of DCBHs. 

We have presumed here that these halos exposed to the intense UV flux by PopII stars are metal-free and remain pristine throughout their evolution. The transition from PopIII to PopII stars may inject metals and pollute these halos. Depending on the critical value of metallicity which can be as low as $\rm10^{-5}$ Z/Z$_{\odot}$ \citep{2009A&A...496..365C,2011ApJ...737...63A,2012A&A...540A.101L}, once the metal content in halos exceeds this value fragmentation becomes inevitable \citep{2008ApJ...686..801O}. Particularly, the cooling due to the dust even in the presence of a strong UV flux becomes effective at densities around $\rm 10^{12}-10^{15}~cm^{-3}$ and may lead to the formation of dense stellar clusters \citep{2009ApJ...694..302D}. Nevertheless, metal enrichment in the universe is expected to be patchy and pristine halos may exist down to $\rm z>6$.

Given the strong dependence of $\rm J_{21}^{crit}$ on the local thermal conditions as argued above, the local heating/cooling effects such as heating by ambiploar diffusion, turbulence dissipation as well as cooling by HD molecules may change the critical value of the flux by a factor of few. In fact the impact of turbulence and magnetic field in the presence of UV flux was explored by \cite{2013A&A...553L...9V} and they found that in turbulent halos with stronger initial seed fields the value of $\rm J_{21}^{crit}$ is reduced by an order of magnitude. Furthermore, the presence of cosmic rays/X-rays may significantly enhance the critical value of $\rm J_{21}$ \citep{2011MNRAS.416.2748I}. It was recently pointed out by \cite{2014arXiv1403.6155R} that including the effect of turbulence in Doppler broadening reduces the $\rm H_{2}$ self-shielding. This should be explored in future studies.

\section*{Acknowledgments}
The simulations described in this work were performed using the Enzo code, developed by the Laboratory for Computational Astrophysics at the University of California in San Diego (http://lca.ucsd.edu). We thank Zoltan Haiman and Greg Bryan for useful discussions on the topic. We acknowledge research funding by Deutsche Forschungsgemeinschaft (DFG) under grant SFB $\rm 963/1$ (project A12) and computing time from HLRN under project nip00029. DRGS and SB thank the DFG for funding via the Schwerpunktprogram SPP 1573 ``Physics of the Interstellar Medium'' (grant SCHL $\rm 1964/1-1$). The simulation results are analyzed using the visualization toolkit for astrophysical data YT \citep{2011ApJS..192....9T}.

\bibliography{blackholes.bib}

\begin{thebibliography}{}

\bibitem[\protect\citeauthoryear{{Agarwal}, {Dalla Vecchia}, {Johnson},
  {Khochfar} \& {Paardekooper}}{{Agarwal} et~al.}{2014}]{2014arXiv1403.5267A}
{Agarwal} B.,  {Dalla Vecchia} C.,  {Johnson} J.~L.,  {Khochfar} S.,
  {Paardekooper} J.-P.,  2014, ArXiv e-prints:1403.5267

\bibitem[\protect\citeauthoryear{{Agarwal}, {Khochfar}, {Johnson}, {Neistein},
  {Dalla Vecchia} \& {Livio}}{{Agarwal} et~al.}{2012}]{2012MNRAS.425.2854A}
{Agarwal} B.,  {Khochfar} S.,  {Johnson} J.~L.,  {Neistein} E.,  {Dalla
  Vecchia} C.,    {Livio} M.,  2012, \mnras, 425, 2854

\bibitem[\protect\citeauthoryear{{Aykutalp} \& {Spaans}}{{Aykutalp} \&
  {Spaans}}{2011}]{2011ApJ...737...63A}
{Aykutalp} A.,  {Spaans} M.,  2011, \apj, 737, 63

\bibitem[\protect\citeauthoryear{{Aykutalp}, {Wise}, {Meijerink} \&
  {Spaans}}{{Aykutalp} et~al.}{2013}]{2013ApJ...771...50A}
{Aykutalp} A.,  {Wise} J.~H.,  {Meijerink} R.,    {Spaans} M.,  2013, \apj,
  771, 50

\bibitem[\protect\citeauthoryear{{Ball}, {Tout} \& {{\.Z}ytkow}}{{Ball}
  et~al.}{2012}]{2012MNRAS.421.2713B}
{Ball} W.~H.,  {Tout} C.~A.,    {{\.Z}ytkow} A.~N.,  2012, \mnras, 421, 2713

\bibitem[\protect\citeauthoryear{{Ball}, {Tout}, {{\.Z}ytkow} \&
  {Eldridge}}{{Ball} et~al.}{2011}]{2011MNRAS.414.2751B}
{Ball} W.~H.,  {Tout} C.~A.,  {{\.Z}ytkow} A.~N.,    {Eldridge} J.~J.,  2011,
  \mnras, 414, 2751

\bibitem[\protect\citeauthoryear{{Begelman}}{{Begelman}}{2010}]{2010MNRAS.402..673B}
{Begelman} M.~C.,  2010, \mnras, 402, 673

\bibitem[\protect\citeauthoryear{{Begelman}, {Rossi} \& {Armitage}}{{Begelman}
  et~al.}{2008}]{2008MNRAS.387.1649B}
{Begelman} M.~C.,  {Rossi} E.~M.,    {Armitage} P.~J.,  2008, \mnras, 387, 1649

\bibitem[\protect\citeauthoryear{{Begelman}, {Volonteri} \& {Rees}}{{Begelman}
  et~al.}{2006}]{2006MNRAS.370..289B}
{Begelman} M.~C.,  {Volonteri} M.,    {Rees} M.~J.,  2006, \mnras, 370, 289

\bibitem[\protect\citeauthoryear{{Bovino}, {Grassi}, {Latif} \&
  {Schleicher}}{{Bovino} et~al.}{2013}]{2013MNRAS.434L..36B}
{Bovino} S.,  {Grassi} T.,  {Latif} M.~A.,    {Schleicher} D.~R.~G.,  2013,
  \mnras, 434, L36

\bibitem[\protect\citeauthoryear{{Bovino}, {Latif}, {Grassi} \&
  {Schleicher}}{{Bovino} et~al.}{2014}]{2014MNRAS.441.2181B}
{Bovino} S.,  {Latif} M.~A.,  {Grassi} T.,    {Schleicher} D.~R.~G.,  2014,
  \mnras, 441, 2181

\bibitem[\protect\citeauthoryear{{Bromm} \& {Loeb}}{{Bromm} \&
  {Loeb}}{2003a}]{2003ApJ...596...34B}
{Bromm} V.,  {Loeb} A.,  2003a, \apj, 596, 34

\bibitem[\protect\citeauthoryear{{Bromm} \& {Loeb}}{{Bromm} \&
  {Loeb}}{2003b}]{Bromm03}
{Bromm} V.,  {Loeb} A.,  2003b, \apj, 596, 34

\bibitem[\protect\citeauthoryear{{Bryan}, {Norman}, {O'Shea}, {Abel}, {Wise},
  {Turk}, {Reynolds}, {Collins}, {So}, {Zhao}, {Cen}, {Li} \& {The Enzo
  Collaboration}}{{Bryan} et~al.}{2014}]{2014ApJS..211...19B}
{Bryan} G.~L.,  {Norman} M.~L.,  {O'Shea} B.~W.,  {Abel} T.,  {Wise} J.~H.,
  {Turk} M.~J.,  {Reynolds} D.~R.,  {Collins} D.~C.,  {So} G.~C.,  {Zhao} F.,
  {Cen} R.,  {Li} Y.,    {The Enzo Collaboration} 2014, \apjs, 211, 19

\bibitem[\protect\citeauthoryear{{Cazaux} \& {Spaans}}{{Cazaux} \&
  {Spaans}}{2009}]{2009A&A...496..365C}
{Cazaux} S.,  {Spaans} M.,  2009, \aap, 496, 365

\bibitem[\protect\citeauthoryear{{Devecchi} \& {Volonteri}}{{Devecchi} \&
  {Volonteri}}{2009}]{2009ApJ...694..302D}
{Devecchi} B.,  {Volonteri} M.,  2009, \apj, 694, 302

\bibitem[\protect\citeauthoryear{{Dijkstra}, {Haiman}, {Mesinger} \&
  {Wyithe}}{{Dijkstra} et~al.}{2008}]{2008MNRAS.391.1961D}
{Dijkstra} M.,  {Haiman} Z.,  {Mesinger} A.,    {Wyithe} J.~S.~B.,  2008,
  \mnras, 391, 1961

\bibitem[\protect\citeauthoryear{{Djorgovski}, {Volonteri}, {Springel}, {Bromm}
  \& {Meylan}}{{Djorgovski} et~al.}{2008}]{2008arXiv0803.2862D}
{Djorgovski} S.~G.,  {Volonteri} M.,  {Springel} V.,  {Bromm} V.,    {Meylan}
  G.,  2008, ArXiv e-prints-0803.2862

\bibitem[\protect\citeauthoryear{{Dopcke}, {Glover}, {Clark} \&
  {Klessen}}{{Dopcke} et~al.}{2013}]{2013ApJ...766..103D}
{Dopcke} G.,  {Glover} S.~C.~O.,  {Clark} P.~C.,    {Klessen} R.~S.,  2013,
  \apj, 766, 103

\bibitem[\protect\citeauthoryear{{Draine} \& {Bertoldi}}{{Draine} \&
  {Bertoldi}}{1996}]{1996ApJ...468..269D}
{Draine} B.~T.,  {Bertoldi} F.,  1996, \apj, 468, 269

\bibitem[\protect\citeauthoryear{{Fan}, {Strauss}, {Richards}, {Hennawi},
  {Becker}, {White} \& {Diamond-Stanic}}{{Fan}
  et~al.}{2006}]{2006AJ....131.1203F}
{Fan} X.,  {Strauss} M.~A.,  {Richards} G.~T.,  {Hennawi} J.~F.,  {Becker}
  R.~H.,  {White} R.~L.,    {Diamond-Stanic} A.~M.,  2006, \aj, 131, 1203

\bibitem[\protect\citeauthoryear{{Fan}, {Strauss}, {Schneider}, {Becker},
  {White}, {Haiman} \& {Gregg}}{{Fan} et~al.}{2003}]{2003AJ....125.1649F}
{Fan} X.,  {Strauss} M.~A.,  {Schneider} D.~P.,  {Becker} R.~H.,  {White}
  R.~L.,  {Haiman} Z.,    {Gregg} M.,  2003, \aj, 125, 1649

\bibitem[\protect\citeauthoryear{{Galli} \& {Palla}}{{Galli} \&
  {Palla}}{1998}]{1998A&A...335..403G}
{Galli} D.,  {Palla} F.,  1998, \aap, 335, 403

\bibitem[\protect\citeauthoryear{{Glover} \& {Abel}}{{Glover} \&
  {Abel}}{2008}]{2008MNRAS.388.1627G}
{Glover} S.~C.~O.,  {Abel} T.,  2008, \mnras, 388, 1627

\bibitem[\protect\citeauthoryear{{Grassi}, {Bovino}, {Schleicher}, {Prieto},
  {Seifried}, {Simoncini} \& {Gianturco}}{{Grassi}
  et~al.}{2014}]{2014MNRAS.439.2386G}
{Grassi} T.,  {Bovino} S.,  {Schleicher} D.~R.~G.,  {Prieto} J.,  {Seifried}
  D.,  {Simoncini} E.,    {Gianturco} F.~A.,  2014, \mnras, 439, 2386

\bibitem[\protect\citeauthoryear{{Haiman}}{{Haiman}}{2004}]{2004ApJ...613...36H}
{Haiman} Z.,  2004, \apj, 613, 36

\bibitem[\protect\citeauthoryear{{Haiman}}{{Haiman}}{2012}]{2012arXiv1203.6075H}
{Haiman} Z.,  2012, ArXiv e-prints-1203.6075

\bibitem[\protect\citeauthoryear{{Haiman} \& {Loeb}}{{Haiman} \&
  {Loeb}}{2001}]{2001ApJ...552..459H}
{Haiman} Z.,  {Loeb} A.,  2001, \apj, 552, 459

\bibitem[\protect\citeauthoryear{{Hirano}, {Hosokawa}, {Yoshida}, {Umeda},
  {Omukai}, {Chiaki} \& {Yorke}}{{Hirano} et~al.}{2014}]{2014ApJ...781...60H}
{Hirano} S.,  {Hosokawa} T.,  {Yoshida} N.,  {Umeda} H.,  {Omukai} K.,
  {Chiaki} G.,    {Yorke} H.~W.,  2014, \apj, 781, 60

\bibitem[\protect\citeauthoryear{{Hosokawa}, {Omukai} \& {Yorke}}{{Hosokawa}
  et~al.}{2012}]{2012ApJ...756...93H}
{Hosokawa} T.,  {Omukai} K.,    {Yorke} H.~W.,  2012, \apj, 756, 93

\bibitem[\protect\citeauthoryear{{Hosokawa}, {Yorke}, {Inayoshi}, {Omukai} \&
  {Yoshida}}{{Hosokawa} et~al.}{2013}]{2013ApJ...778..178H}
{Hosokawa} T.,  {Yorke} H.~W.,  {Inayoshi} K.,  {Omukai} K.,    {Yoshida} N.,
  2013, \apj, 778, 178

\bibitem[\protect\citeauthoryear{{Inayoshi} \& {Omukai}}{{Inayoshi} \&
  {Omukai}}{2011}]{2011MNRAS.416.2748I}
{Inayoshi} K.,  {Omukai} K.,  2011, \mnras, 416, 2748

\bibitem[\protect\citeauthoryear{{Inayoshi}, {Omukai} \& {Tasker}}{{Inayoshi}
  et~al.}{2014}]{2014arXiv1404.4630I}
{Inayoshi} K.,  {Omukai} K.,    {Tasker} E.~J.,  2014, ArXiv e-prints:1404.4630

\bibitem[\protect\citeauthoryear{{Johnson}, {Khochfar}, {Greif} \&
  {Durier}}{{Johnson} et~al.}{2010}]{2010MNRAS.tmp.1427J}
{Johnson} J.~L.,  {Khochfar} S.,  {Greif} T.~H.,    {Durier} F.,  2010, \mnras,
  pp 1427--+

\bibitem[\protect\citeauthoryear{{Latif}, {Niemeyer} \& {Schleicher}}{{Latif}
  et~al.}{2014}]{2014MNRAS.440.2969L}
{Latif} M.~A.,  {Niemeyer} J.~C.,    {Schleicher} D.~R.~G.,  2014, \mnras, 440,
  2969

\bibitem[\protect\citeauthoryear{{Latif}, {Schleicher} \& {Schmidt}}{{Latif}
  et~al.}{2014}]{2014MNRAS.tmp..564L}
{Latif} M.~A.,  {Schleicher} D.~R.~G.,    {Schmidt} W.,  2014, \mnras

\bibitem[\protect\citeauthoryear{{Latif}, {Schleicher}, {Schmidt} \&
  {Niemeyer}}{{Latif} et~al.}{2013a}]{2013MNRAS.433.1607L}
{Latif} M.~A.,  {Schleicher} D.~R.~G.,  {Schmidt} W.,    {Niemeyer} J.,  2013a,
  \mnras, 433, 1607

\bibitem[\protect\citeauthoryear{{Latif}, {Schleicher}, {Schmidt} \&
  {Niemeyer}}{{Latif} et~al.}{2013b}]{2013MNRAS.430..588L}
{Latif} M.~A.,  {Schleicher} D.~R.~G.,  {Schmidt} W.,    {Niemeyer} J.,  2013b,
  \mnras, 430, 588

\bibitem[\protect\citeauthoryear{{Latif}, {Schleicher}, {Schmidt} \&
  {Niemeyer}}{{Latif} et~al.}{2013c}]{2013MNRAS.432..668L}
{Latif} M.~A.,  {Schleicher} D.~R.~G.,  {Schmidt} W.,    {Niemeyer} J.,  2013c,
  \mnras, 432, 668

\bibitem[\protect\citeauthoryear{{Latif}, {Schleicher}, {Schmidt} \&
  {Niemeyer}}{{Latif} et~al.}{2013d}]{2013MNRAS.tmp.2526L}
{Latif} M.~A.,  {Schleicher} D.~R.~G.,  {Schmidt} W.,    {Niemeyer} J.~C.,
  2013d, \mnras

\bibitem[\protect\citeauthoryear{{Latif}, {Schleicher}, {Schmidt} \&
  {Niemeyer}}{{Latif} et~al.}{2013e}]{2013MNRAS.436.2989L}
{Latif} M.~A.,  {Schleicher} D.~R.~G.,  {Schmidt} W.,    {Niemeyer} J.~C.,
  2013e, \mnras, 436, 2989

\bibitem[\protect\citeauthoryear{{Latif}, {Schleicher} \& {Spaans}}{{Latif}
  et~al.}{2012}]{2012A&A...540A.101L}
{Latif} M.~A.,  {Schleicher} D.~R.~G.,    {Spaans} M.,  2012, \aap, 540, A101

\bibitem[\protect\citeauthoryear{{Latif}, {Zaroubi} \& {Spaans}}{{Latif}
  et~al.}{2011}]{2011MNRAS.411.1659L}
{Latif} M.~A.,  {Zaroubi} S.,    {Spaans} M.,  2011, \mnras, 411, 1659

\bibitem[\protect\citeauthoryear{{Lodato} \& {Natarajan}}{{Lodato} \&
  {Natarajan}}{2006}]{2006MNRAS.371.1813L}
{Lodato} G.,  {Natarajan} P.,  2006, \mnras, 371, 1813

\bibitem[\protect\citeauthoryear{{Madau}, {Haardt} \& {Dotti}}{{Madau}
  et~al.}{2014}]{2014ApJ...784L..38M}
{Madau} P.,  {Haardt} F.,    {Dotti} M.,  2014, \apjl, 784, L38

\bibitem[\protect\citeauthoryear{{Martin}, {Schwarz} \& {Mandy}}{{Martin}
  et~al.}{1996}]{1996ApJ...461..265M}
{Martin} P.~G.,  {Schwarz} D.~H.,    {Mandy} M.~E.,  1996, \apj, 461, 265

\bibitem[\protect\citeauthoryear{{Mortlock}, {Warren}, {Venemans}, {Patel},
  {Hewett}, {McMahon}, {Simpson}, {Theuns}, {Gonz{\'a}les-Solares}, {Adamson},
  {Dye}, {Hambly}, {Hirst}, {Irwin}, {Kuiper}, {Lawrence} \&
  {R{\"o}ttgering}}{{Mortlock} et~al.}{2011}]{2011Natur.474..616M}
{Mortlock} D.~J.,  {Warren} S.~J.,  {Venemans} B.~P.,  {Patel} M.,  {Hewett}
  P.~C.,  {McMahon} R.~G.,  {Simpson} C.,  {Theuns} T.,  {Gonz{\'a}les-Solares}
  E.~A.,  {Adamson} A.,  {Dye} S.,  {Hambly} N.~C.,  {Hirst} P.,  {Irwin}
  M.~J.,  {Kuiper} E.,  {Lawrence} A.,    {R{\"o}ttgering} H.~J.~A.,  2011,
  \nat, 474, 616

\bibitem[\protect\citeauthoryear{{Oh} \& {Haiman}}{{Oh} \&
  {Haiman}}{2002}]{2002ApJ...569..558O}
{Oh} S.~P.,  {Haiman} Z.,  2002, \apj, 569, 558

\bibitem[\protect\citeauthoryear{{Omukai}}{{Omukai}}{2001}]{2001ApJ...546..635O}
{Omukai} K.,  2001, \apj, 546, 635

\bibitem[\protect\citeauthoryear{{Omukai}, {Schneider} \& {Haiman}}{{Omukai}
  et~al.}{2008}]{2008ApJ...686..801O}
{Omukai} K.,  {Schneider} R.,    {Haiman} Z.,  2008, \apj, 686, 801

\bibitem[\protect\citeauthoryear{{Omukai}, {Tsuribe}, {Schneider} \&
  {Ferrara}}{{Omukai} et~al.}{2005}]{2005ApJ...626..627O}
{Omukai} K.,  {Tsuribe} T.,  {Schneider} R.,    {Ferrara} A.,  2005, \apj, 626,
  627

\bibitem[\protect\citeauthoryear{{O'Shea}, {Bryan}, {Bordner}, {Norman},
  {Abel}, {Harkness} \& {Kritsuk}}{{O'Shea} et~al.}{2004}]{2004astro.ph..3044O}
{O'Shea} B.~W.,  {Bryan} G.,  {Bordner} J.,  {Norman} M.~L.,  {Abel} T.,
  {Harkness} R.,    {Kritsuk} A.,  2004, ArXiv Astrophysics e-prints 0403044

\bibitem[\protect\citeauthoryear{{Portegies Zwart}, {Baumgardt}, {Hut},
  {Makino} \& {McMillan}}{{Portegies Zwart} et~al.}{2004}]{2004Natur.428..724P}
{Portegies Zwart} S.~F.,  {Baumgardt} H.,  {Hut} P.,  {Makino} J.,
  {McMillan} S.~L.~W.,  2004, \nat, 428, 724

\bibitem[\protect\citeauthoryear{{Prieto}, {Jimenez} \& {Haiman}}{{Prieto}
  et~al.}{2013}]{2013MNRAS.436.2301P}
{Prieto} J.,  {Jimenez} R.,    {Haiman} Z.,  2013, \mnras, 436, 2301

\bibitem[\protect\citeauthoryear{{Regan} \& {Haehnelt}}{{Regan} \&
  {Haehnelt}}{2009}]{2009MNRAS.393..858R}
{Regan} J.~A.,  {Haehnelt} M.~G.,  2009, \mnras, 393, 858

\bibitem[\protect\citeauthoryear{{Regan}, {Johansson} \& {Haehnelt}}{{Regan}
  et~al.}{2014}]{2014MNRAS.439.1160R}
{Regan} J.~A.,  {Johansson} P.~H.,    {Haehnelt} M.~G.,  2014, \mnras, 439,
  1160

\bibitem[\protect\citeauthoryear{{Richings}, {Schaye} \&
  {Oppenheimer}}{{Richings} et~al.}{2014}]{2014arXiv1403.6155R}
{Richings} A.~J.,  {Schaye} J.,    {Oppenheimer} B.~D.,  2014, ArXiv
  e-prints:1403.6155

\bibitem[\protect\citeauthoryear{{Schleicher}, {Palla}, {Ferrara}, {Galli} \&
  {Latif}}{{Schleicher} et~al.}{2013}]{2013A&A...558A..59S}
{Schleicher} D.~R.~G.,  {Palla} F.,  {Ferrara} A.,  {Galli} D.,    {Latif} M.,
  2013, \aap, 558, A59

\bibitem[\protect\citeauthoryear{{Schleicher}, {Spaans} \&
  {Glover}}{{Schleicher} et~al.}{2010}]{2010ApJ...712L..69S}
{Schleicher} D.~R.~G.,  {Spaans} M.,    {Glover} S.~C.~O.,  2010, \apjl, 712,
  L69

\bibitem[\protect\citeauthoryear{{Shang}, {Bryan} \& {Haiman}}{{Shang}
  et~al.}{2010}]{2010MNRAS.402.1249S}
{Shang} C.,  {Bryan} G.~L.,    {Haiman} Z.,  2010, \mnras, 402, 1249

\bibitem[\protect\citeauthoryear{{Spaans} \& {Silk}}{{Spaans} \&
  {Silk}}{2006}]{2006ApJ...652..902S}
{Spaans} M.,  {Silk} J.,  2006, \apj, 652, 902

\bibitem[\protect\citeauthoryear{{Tanaka} \& {Haiman}}{{Tanaka} \&
  {Haiman}}{2009}]{2009ApJ...696.1798T}
{Tanaka} T.,  {Haiman} Z.,  2009, \apj, 696, 1798

\bibitem[\protect\citeauthoryear{{Turk}, {Oishi}, {Abel} \& {Bryan}}{{Turk}
  et~al.}{2012}]{2012ApJ...745..154T}
{Turk} M.~J.,  {Oishi} J.~S.,  {Abel} T.,    {Bryan} G.~L.,  2012, \apj, 745,
  154

\bibitem[\protect\citeauthoryear{{Turk}, {Smith}, {Oishi}, {Skory}, {Skillman},
  {Abel} \& {Norman}}{{Turk} et~al.}{2011}]{2011ApJS..192....9T}
{Turk} M.~J.,  {Smith} B.~D.,  {Oishi} J.~S.,  {Skory} S.,  {Skillman} S.~W.,
  {Abel} T.,    {Norman} M.~L.,  2011, \apjs, 192, 9

\bibitem[\protect\citeauthoryear{{Van Borm} \& {Spaans}}{{Van Borm} \&
  {Spaans}}{2013}]{2013A&A...553L...9V}
{Van Borm} C.,  {Spaans} M.,  2013, \aap, 553, L9

\bibitem[\protect\citeauthoryear{{Venemans}, {Findlay}, {Sutherland}, {De
  Rosa}, {McMahon}, {Simcoe}, {Gonz{\'a}lez-Solares}, {Kuijken} \&
  {Lewis}}{{Venemans} et~al.}{2013}]{2013ApJ...779...24V}
{Venemans} B.~P.,  {Findlay} J.~R.,  {Sutherland} W.~J.,  {De Rosa} G.,
  {McMahon} R.~G.,  {Simcoe} R.,  {Gonz{\'a}lez-Solares} E.~A.,  {Kuijken} K.,
    {Lewis} J.~R.,  2013, \apj, 779, 24

\bibitem[\protect\citeauthoryear{{Visbal}, {Haiman} \& {Bryan}}{{Visbal}
  et~al.}{2014}]{2014arXiv1403.1293V}
{Visbal} E.,  {Haiman} Z.,    {Bryan} G.~L.,  2014, ArXiv e-prints:1403.1293

\bibitem[\protect\citeauthoryear{{Volonteri}}{{Volonteri}}{2010}]{2010A&ARv..18..279V}
{Volonteri} M.,  2010, \aapr, 18, 279

\bibitem[\protect\citeauthoryear{{Volonteri} \& {Bellovary}}{{Volonteri} \&
  {Bellovary}}{2012}]{2012RPPh...75l4901V}
{Volonteri} M.,  {Bellovary} J.,  2012, Reports on Progress in Physics, 75,
  124901

\bibitem[\protect\citeauthoryear{{Whalen}, {Even}, {Lovekin}, {Fryer},
  {Stiavelli}, {Roming}, {Cooke}, {Pritchard}, {Holz} \& {Knight}}{{Whalen}
  et~al.}{2013}]{2013ApJ...768..195W}
{Whalen} D.~J.,  {Even} W.,  {Lovekin} C.~C.,  {Fryer} C.~L.,  {Stiavelli} M.,
  {Roming} P.~W.~A.,  {Cooke} J.,  {Pritchard} T.~A.,  {Holz} D.~E.,
  {Knight} C.,  2013, \apj, 768, 195

\bibitem[\protect\citeauthoryear{{Whalen} \& {Fryer}}{{Whalen} \&
  {Fryer}}{2012}]{2012ApJ...756L..19W}
{Whalen} D.~J.,  {Fryer} C.~L.,  2012, \apjl, 756, L19

\bibitem[\protect\citeauthoryear{{Whalen}, {Johnson}, {Smidt}, {Meiksin},
  {Heger}, {Even} \& {Fryer}}{{Whalen} et~al.}{2013}]{2013ApJ...774...64W}
{Whalen} D.~J.,  {Johnson} J.~L.,  {Smidt} J.,  {Meiksin} A.,  {Heger} A.,
  {Even} W.,    {Fryer} C.~L.,  2013, \apj, 774, 64

\bibitem[\protect\citeauthoryear{{Willott}, {Delorme}, {Reyl{\'e}}, {Albert},
  {Bergeron}, {Crampton}, {Delfosse}, {Forveille}, {Hutchings}, {McLure},
  {Omont} \& {Schade}}{{Willott} et~al.}{2010}]{2010AJ....139..906W}
{Willott} C.~J.,  {Delorme} P.,  {Reyl{\'e}} C.,  {Albert} L.,  {Bergeron} J.,
  {Crampton} D.,  {Delfosse} X.,  {Forveille} T.,  {Hutchings} J.~B.,  {McLure}
  R.~J.,  {Omont} A.,    {Schade} D.,  2010, \aj, 139, 906

\bibitem[\protect\citeauthoryear{{Wise}, {Turk} \& {Abel}}{{Wise}
  et~al.}{2008}]{2008ApJ...682..745W}
{Wise} J.~H.,  {Turk} M.~J.,    {Abel} T.,  2008, \apj, 682, 745

\bibitem[\protect\citeauthoryear{{Wolcott-Green}, {Haiman} \&
  {Bryan}}{{Wolcott-Green} et~al.}{2011}]{2011MNRAS.418..838W}
{Wolcott-Green} J.,  {Haiman} Z.,    {Bryan} G.~L.,  2011, \mnras, 418, 838

\bibitem[\protect\citeauthoryear{{Wutschik}, {Schleicher} \&
  {Palmer}}{{Wutschik} et~al.}{2013}]{2013A&A...560A..34W}
{Wutschik} S.,  {Schleicher} D.~R.~G.,    {Palmer} T.~S.,  2013, \aap, 560, A34

\bibitem[\protect\citeauthoryear{{Yue}, {Ferrara}, {Salvaterra}, {Xu} \&
  {Chen}}{{Yue} et~al.}{2013}]{2013MNRAS.433.1556Y}
{Yue} B.,  {Ferrara} A.,  {Salvaterra} R.,  {Xu} Y.,    {Chen} X.,  2013,
  \mnras, 433, 1556

\bibitem[\protect\citeauthoryear{{Yue}, {Ferrara}, {Salvaterra}, {Xu} \&
  {Chen}}{{Yue} et~al.}{2014}]{2014MNRAS.440.1263Y}
{Yue} B.,  {Ferrara} A.,  {Salvaterra} R.,  {Xu} Y.,    {Chen} X.,  2014,
  \mnras, 440, 1263

\end{thebibliography}

\end{document}